\documentclass{mn2e}
\def\spose#1{\hbox to 0pt{#1\hss}}
\def\approxgt{\mathrel{\spose{\lower 3pt\hbox{$\sim$}}\raise 2.0pt\hbox{$>$}}}
\def\approxlt{\mathrel{\spose{\lower 3pt\hbox{$\sim$}}\raise 2.0pt\hbox{$<$}}}
\def\etal{{\it et al.\ }}
\def\eg{{\it e.g.\ }}
\usepackage{color,hyperref,graphics,graphicx}

\title[Ram pressure stripping in the Ophiuchus Cluster]
{Ram pressure stripping of the cool core of 
the Ophiuchus Cluster}
\author[E. T. Million et al.]{
E.~T.~Million,$^1$ S.~W.~Allen,$^1$ N.~Werner$^1$ and G.~B.~Taylor$^2$ \\
\footnotesize
  $^1$Kavli Institute for Particle Astrophysics and Cosmology, Stanford University, 382 Via Pueblo Mall, Stanford, CA 94305-4060, USA; \\
SLAC National Accelerator Laboratory, 2575 Sand Hill Road, Menlo Park, CA 94025, USA\\
  $^2$Department of Physics and Astronomy, University of New Mexico, Albuquerque NM, 87131, USA\\
}
\begin{document}
\renewcommand{\thefootnote}{\arabic{\footnote}}
\maketitle
\begin{abstract}
We report results from a $Chandra$ study of the central regions of the
nearby, X-ray bright, Ophiuchus Cluster ($z=0.03$),
the second-brightest cluster in the sky.
Our study reveals
a dramatic, close-up view of the stripping and potential destruction
of a cool core within a rich cluster.
The X-ray emission from the Ophiuchus Cluster core exhibits a
comet-like morphology extending to the north, driven by merging activity, 
indicative of ram-pressure stripping caused by
rapid motion through the ambient cluster gas.  A cold front at the
southern edge implies a velocity of $1000\pm200$ km s$^{-1}$ (M$\sim$0.6).
The X-ray
emission from the cluster core is sharply peaked.
As previously noted, the peak is
offset by 4 arcsec ($\sim2$ kpc) from the optical center of the
associated cD galaxy, indicating that ram pressure has slowed the
core, allowing the relatively collisionless stars and dark matter to
carry on ahead.  
The cluster exhibits the strongest central
temperature gradient of any massive cluster observed to date: 
the temperature rises
from 0.7 keV within 1 kpc of the brightness peak, to 10 keV by 30 kpc.
A strong metallicity gradient is also observed within the same region.
This supports a picture in which the outer parts of the cool core have
been stripped by ram-pressure due to its rapid motion. 
The cooling
time of the innermost gas is very short, $\sim5\times10^7$ yrs. 
Within
the central 10 kpc radius, multiple small-scale fronts and a complex
thermodynamic structure are observed, indicating significant 
motions. These may be excited by the separation of the X-ray peak from
the cD galaxy and its associated dark matter potential.  
Beyond the
central 50 kpc, and out to a radius $\sim150$ kpc, the cluster appears
relatively isothermal and has near constant metallicity.
The exception is a large, coherent ridge of enhanced metallicity observed
to trail the cool core, and which is likely to have been stripped from it.
\end{abstract}


\section{Introduction}

\begin{figure*}
\hspace{0.0cm}
\scalebox{0.43}{\includegraphics{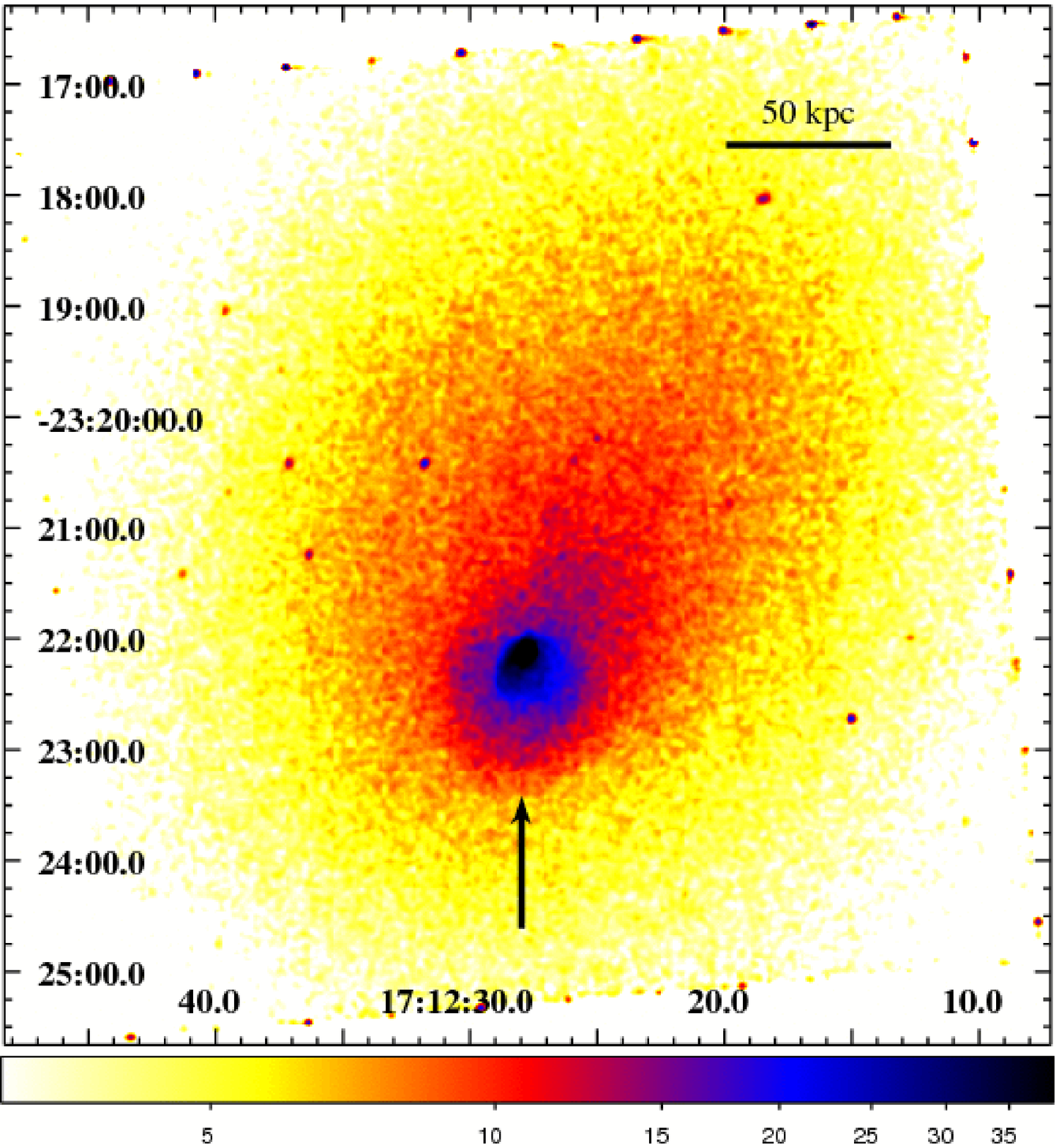}}
\hspace{0.9cm}
\scalebox{0.43}{\includegraphics{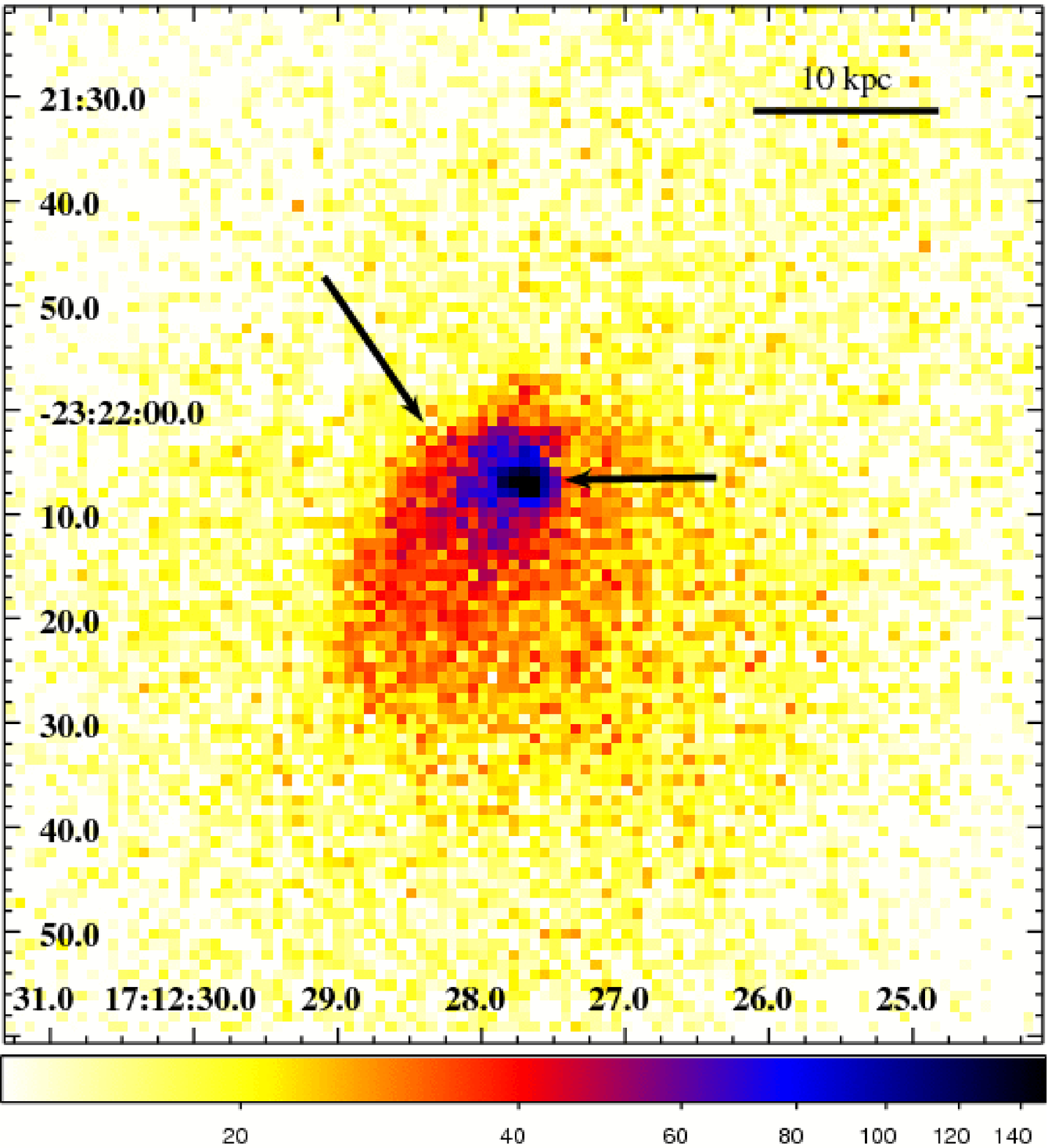}}
\caption{Background subtracted, flat-fielded X-ray surface
brightness maps in the $0.6-7.0$ keV band, smoothed with a 2 arcsec
Gaussian filter.  
The three surface brightness discontinuities noted by 
Ascasibar \& Markevitch (2006) are marked with arrows.
(a) The central $8\times8$ arcmin$^2$ region
covered by ACIS chip 7. 
Note the comet-like extension to the north/northwest and the cold 
front observed $\sim$40 kpc to the south of the X-ray peak (see 
also Ascasibar \& Markevitch 2006).
(b) The
central $1.6\times1.6$ arcmin$^2$ region of the cluster.  Two additional
surface brightness discontinuities are observed.  One is
10 arcsec (5.5 kpc) to the northeast of the X-ray peak, the other is 3
arcsec (1.5 kpc) to the west (see also Ascasibar \& Markevitch 2006).
}
\label{fig:sb}
\end{figure*}

The Ophiuchus Cluster ($z=0.03$) is the second-brightest cluster of
galaxies in the $2-10$ keV sky (Edge \etal 1990; see also Ebeling
\etal 2002).  This extremely massive, nearby system lies behind the
plane of our Galaxy and is highly obscured, with a line-of-sight
equivalent hydrogen column density $N_H\sim3\times10^{21}$ atom
cm$^{-2}$.  Using data from the Einstein Observatory and EXOSAT,
Arnaud \etal (1987) showed that the cluster has a high mean
temperature, $kT=9.4^{+1.5}_{-1.2}$ keV (90 per cent confidence
errors) and hosts a sharply-peaked cooling core.
These findings were
confirmed by later studies with ASCA (Matsuzawa \etal 1996, Watanabe
\etal 2001) and $Chandra$ (Sun \etal 2007).  Relatively cool, X-ray
bright spectral components are associated with the cluster center,
with a minimum measured temperature below 1 keV (P\'erez-Torres
\etal 2009).
The peak of the X-ray emission from the cluster is also offset
from the X-ray centroid at larger scales (Arnaud \etal 1987).

Ascasibar \& Markevitch (2006) noted the presence of
three surface brightness discontinuities in the central regions
of the Ophiuchus Cluster (see also Fig. \ref{fig:sb}),
suggesting significant
motion of the cool, dense core. Such fronts are common in clusters
and are typically caused by 
merger events (\eg Vikhlinin, Markevitch \& Murray
2001; Reiprich \etal 2001). However, many relatively relaxed clusters, such as 
Abell 1795 and 2029 (Fabian \etal 2001; Markevitch, Vikhlinin \& Mazzotta 2001; 
Clarke \etal 2004), also exhibit
surface brightness discontinuities near their X-ray surface brightness
peaks. Simulations suggest that `sloshing' of the gas even in cool
core clusters can lead to visible fronts up to several Gyrs after a 
relatively minor merger event
(see Tittley \& Henriksen 2005; Ascasibar \& Markevitch 2006; see also 
Markevitch \& Vikhlinin 2007 and references within). A cool core in motion
within the ambient ICM 
will also experience ram pressure. The impact of this ram pressure
is of particular interest for the survivability of cool cores following
major mergers.

Due to its location and large angular size, optical studies of the
Ophiuchus Cluster have proved challenging.  However, detailed 
spectroscopic studies
suggest that the system is very massive, with a velocity
dispersion $\sigma_v=1050\pm50$ km s$^{-1}$ (Wakamatsu et al. 2005),
consistent with the measured X-ray temperature and luminosity (Edge
\etal 1990).  Clear substructure in the galaxy velocity histogram is
observed, indicating that the cluster is merging (Wakamatsu \etal
2005).  
The dominant cluster galaxy exhibits a substantial peculiar
velocity, $v_p=$-650 km\,s$^{-1}$, with respect to the cluster mean,
$cz=9045\pm30$ km\,s$^{-1}$ (Wakamatsu \etal 2005; Hasegawa \etal
2000).  The presence of many nearby, smaller clusters and groups
argues that Ophiuchus lies at the center of a supercluster
(Wakamatsu \etal 2005; see also Hasegawa \etal 2000).

Govoni \etal (2009; see also Murgia \etal 2009) report the discovery
of a low surface brightness, diffuse radio mini-halo surrounding the
dominant cluster galaxy, 2MASX\,J17122774-2322108.  
The central cD galaxy is also detected at 1.4 GHz 
with a flux density of 29.3 mJy in the NVSS
survey (Condon \etal 1998; see also Section \ref{section:image}) with
a $45\times45$ arcsec$^2$ (FWHM) synthesized beam.
There are a few other strong radio sources at the periphery
of the cluster (see Johnston \etal 1981).
However, their association with the Ophiuchus Cluster
is uncertain. A variety of
authors have searched for non-thermal X-ray emission from the cluster
in the hard X-ray band (see Nevalainen \etal 2004, 2009; 
Eckert \etal 2008;
see also Dunn \& Fabian 2006), although the results to date remain
inconclusive (see Fujita \etal 2008; Ajello \etal 2009).

In this paper, we report detailed $Chandra$ observations of the core of
Ophiuchus Cluster. Section 2 discusses the data reduction and analysis
method. Section 3 describes the main results and Section 4 provides 
physical interpretation.  At a redshift of $z=0.03$, one arcsec
corresponds to 0.60 kpc, assuming a $\Lambda$CDM cosmology of
$H_0=70$ km s$^{-1}$ Mpc$^{-1}$, $\Omega_M=0.3$, and
$\Omega_\Lambda=0.7$.

\section{Observations and Analysis}

\subsection{X-ray data reduction}

The $Chandra$ observation of the Ophiuchus Cluster (observation ID 3200)
was carried out using the Advanced CCD Imaging Spectrometer (ACIS) in
October 2002.  The standard level-1 event lists produced by the
$Chandra$ pipeline were reprocessed using the $CIAO$ (version 4.1.2)
software package, including the appropriate gain maps and calibration
products ($CALDB$ version 4.1.3).  Bad pixels were removed and
standard grade selections applied.  The data were cleaned to remove
periods of anomalously high background, using the standard energy
ranges and binning methods recommended by the $Chandra$ X-ray
Center. The resulting net, clean exposure time is 49.5 ks. Separate
photon-weighted response matrices and effective area files were
constructed for each region analyzed.

\begin{figure}
\hspace{0.0cm}
\scalebox{0.43}{\includegraphics{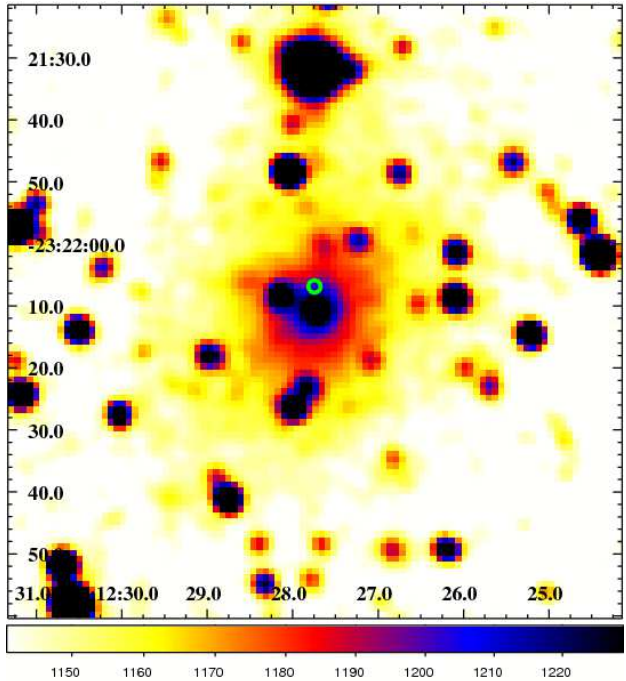}}
\caption{Co-added H, J, and K band 2MASS image of the inner regions of
the Ophiuchus Cluster, showing the same region as Fig. 1b.  The green 
circle
marks the position of the X-ray brightness peak. The radius of the
circle (1 arcsec) represents a conservative estimate of the
astrometric uncertainties. 
As reported by P\'erez-Torres \etal (2009),
the X-ray peak is offset by $\sim$2 kpc (4
arcsec) to the north of the central galaxy.
}
\label{fig:optical}
\end{figure}

\begin{figure}
\hspace{0.0cm}
\scalebox{0.43}{\includegraphics{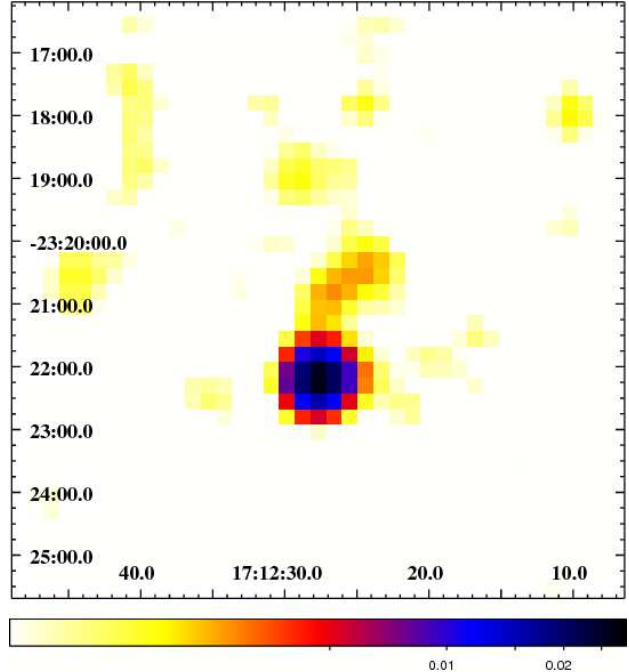}}
\caption{1.4 GHz NVSS image (Condon \etal 1998) of the Ophiuchus Cluster
spanning the same region as figure \ref{fig:sb}a.  
A bright central
radio source is coincident with the location of the central galaxy.  
The spatial resolution of the radio data is approximately $45\times45$
arcsec$^2$ (FWHM; see also P\'erez-Torres \etal 2009).
The radio emission extends to the northwest, similar to the comet-like X-ray
morphology.
}
\label{fig:radio}
\end{figure}

\begin{figure}
\hspace{-0.1cm}
\scalebox{0.47}{\includegraphics[angle=270]{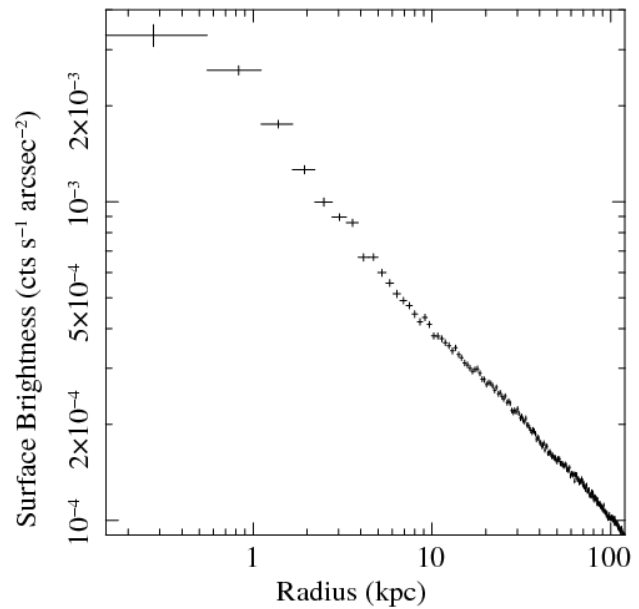}}
\caption{Azimuthally-averaged surface brightness profile 
in the $0.6-7.0$ keV band, centered on the X-ray peak. 
The profile is very sharp and no clear central flattening is observed.
}
\label{fig:sbprof}
\end{figure}

\subsection{Spatially-resolved X-ray spectroscopy}

\subsubsection{Spatial binning}

The individual regions for the spectral analysis were determined using
the contour binning method of Sanders (2006), which groups neighboring
pixels of similar surface brightness until a desired signal-to-noise
threshold is met.  For data of the quality discussed here, the regions
are small enough that the X-ray emission from each can be approximated
usefully by a single temperature plasma model.  We have focused our
analysis on data from the central back-illuminated ACIS-S chip (chip 7).

We initially carried out thermodynamic mapping with $\sim$10,000 net
counts per region, to examine the thermodynamic structure of the 
hottest gas in the cluster on large scales.  This allows us to measure the
metallicity and temperature of gas with kT$\sim10$ keV to an accuracy
of $\sim$10 per cent.  Higher-resolution thermodynamic mapping of the
central 40 kpc was also performed, with $\sim$3,000 net counts per
region, to examine the thermodynamic structure of the cooler cluster
core.  This allows us to measure the metallicity of this gas
to better than 20 per
cent, and the temperature of regions with kT$<5$ keV to $\sim$5 per
cent accuracy or better.

In order to determine radial profiles for thermodynamic properties, we
have also extracted spectra in concentric circular annuli, centered
on the X-ray peak position. 
The widths of the annuli range from 1 arcsec to 40 arcsec.

\subsubsection{Background modelling}

Background spectra for the appropriate detector regions were extracted
from the blank-sky fields available from the $Chandra$ X-ray Center.
The blank-sky fields were processed in an identical manner to the 
Ophiuchus Cluster observation,
and were reprojected onto the same coordinates using the appropriate aspect
solution files.
Background regions were chosen to match the same extraction regions
as the science data.
The background spectra 
were normalized by the ratio of the observed and blank-sky count
rates in the $9.5-12$ keV band
(due to the very small effective area of the $Chandra$ telescope at hard
X-ray energies, there is no significant cluster emission in the $9.5-12$
keV band).
The statistical uncertainty in the
observed $9.5-12$ keV flux is less than 5 per cent.

To enable further refinement of the background models, we have also
extracted spectra from relatively source-free regions of the other
back-illuminated ACIS CCD (chip
5). Although cluster emission is still clearly present on this offset
chip, it is significantly fainter than on the central chip (Chip 7)
and the data from it can be compared to the blank-sky fields to
identify any additional, strong foreground/background emission components.
Such components are commonly detected in the spectra of sources lying
behind the Galactic Plane. We indeed detect excess soft emission with
respect to the blank sky fields in the Ophiuchus data, which is
modelled in the analysis. We note, however, that our results on the
bright cluster core are relatively insensitive to the details of the
background modelling.

\subsubsection{Spectral analysis}
\label{section:model}

The spectra have been analyzed using {\small XSPEC} (version 12.5; Arnaud
1996), the {\small MEKAL} plasma emission code (Kaastra \& Mewe 1993), and the
photoelectric absorption models of Balucinska-Church \& McCammon
(1992).  All spectral fits were carried out in the $0.6-7.0$ keV
energy band. The extended C-statistic available in {\small XSPEC} was used for
all fitting.

The default spectral model applied to each spatial region consists of
an optically-thin, {\small MEKAL} thermal plasma model, at the redshift of the
cluster. The normalization, temperature, and
metallicity\footnote{Cluster abundances are calculated with respect to
Anders \& Grevesse (1989).}  in each spatial region are free
parameters in the fits.  Since the Galactic column density determined
from HI studies is high (Kalberla \etal 2005; see also Dickey \& Lockman
1990), we include this as an additional free parameter in the
analysis, linked to vary in unison across all regions. We have also
examined models where the column density is allowed to vary spatially.

In Section \ref{section:nonthermal}, 
we briefly discuss results from a search for non-thermal
X-ray emission from the cluster. In this case, additional power-law model
components were included in the analysis, following the method 
described by Million \& Allen (2009).

For the determination of deprojected radial profiles from the
azimuthally-averaged spectra, we use the {\small PROJCT} model 
in {\small XSPEC}.

\begin{figure*}
\hspace{0.0cm}
\scalebox{0.43}{\includegraphics{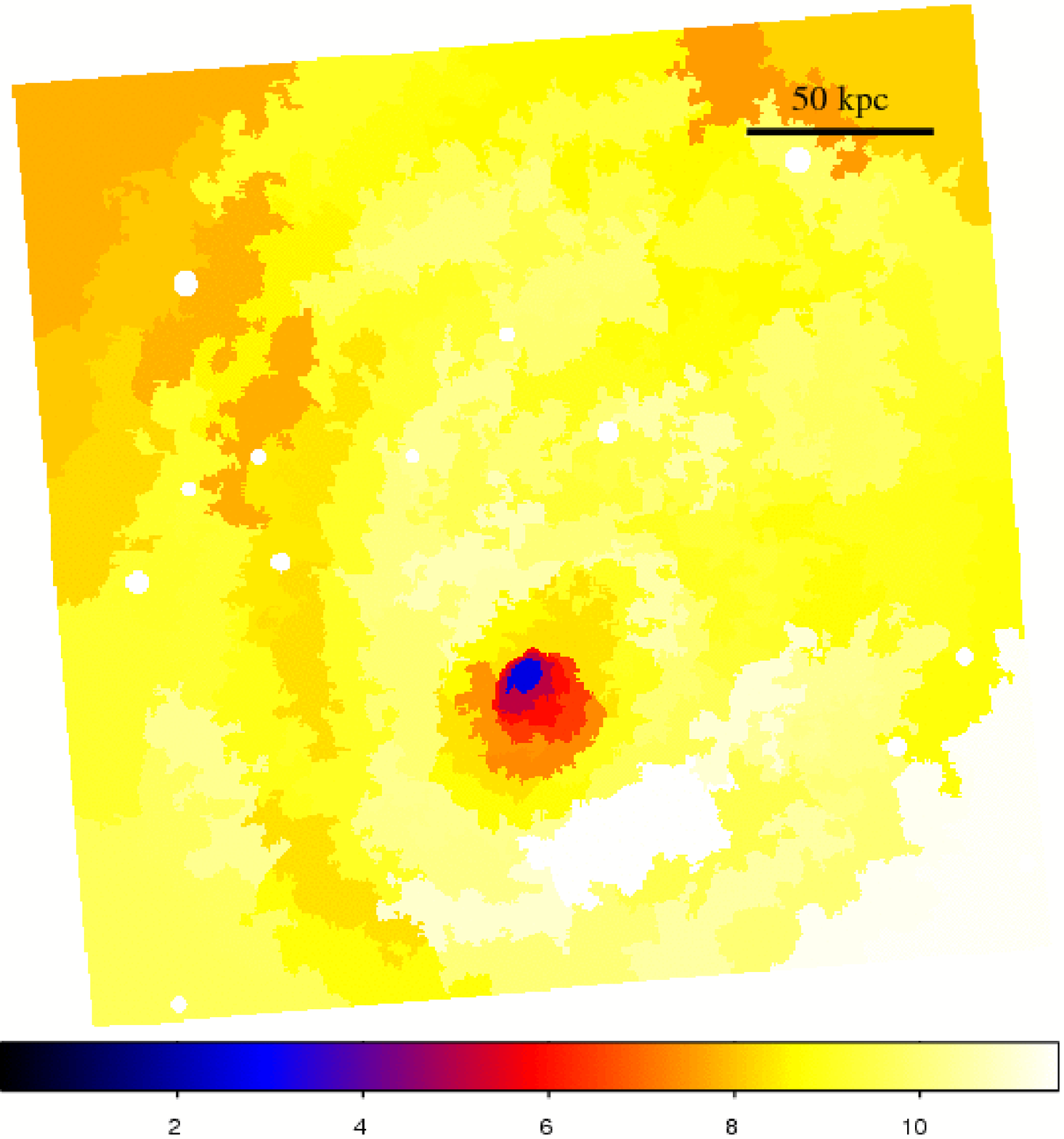}}
\hspace{0.9cm}
\scalebox{0.43}{\includegraphics{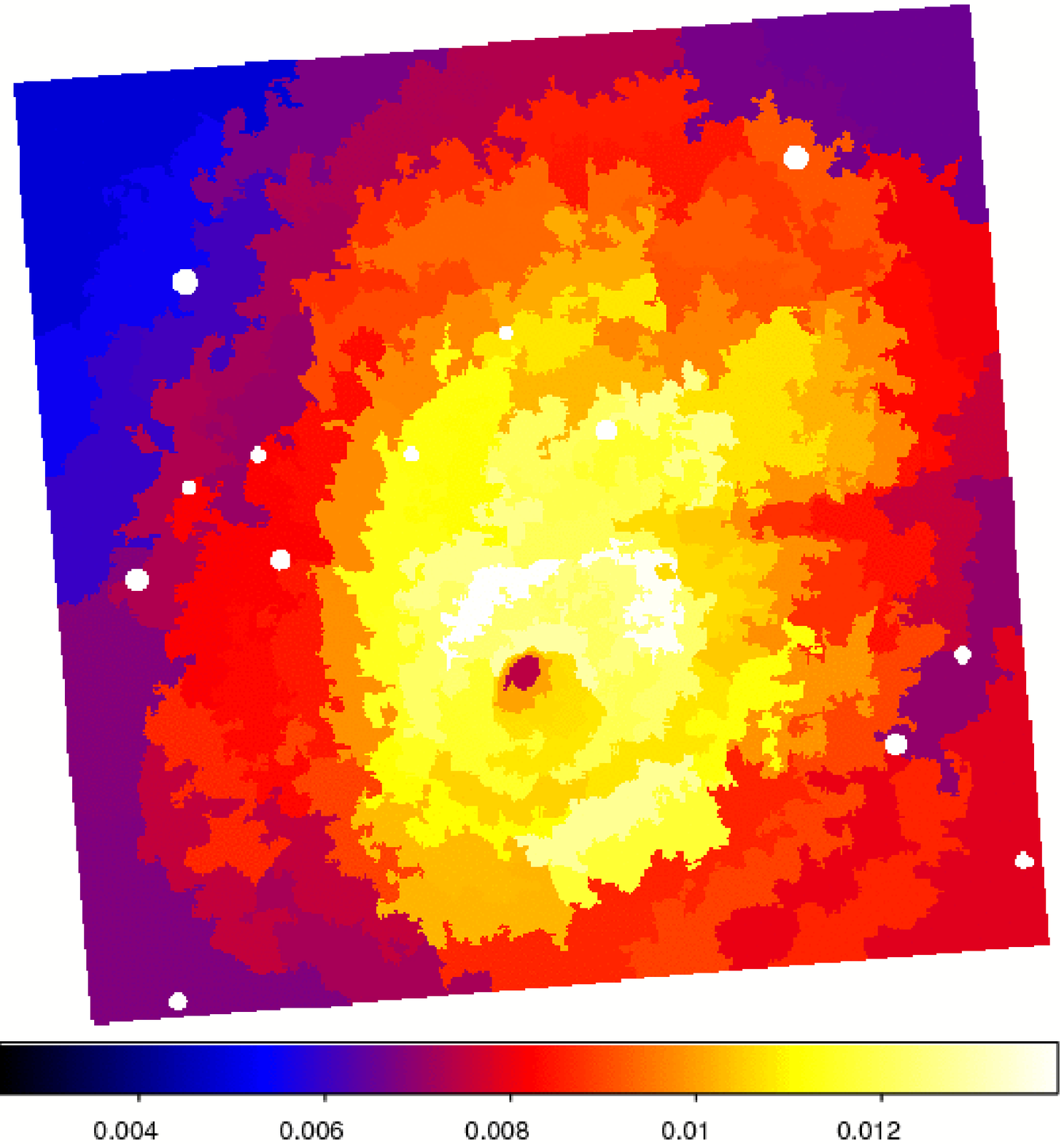}}\\
\hspace{0.0cm}
\scalebox{0.43}{\includegraphics{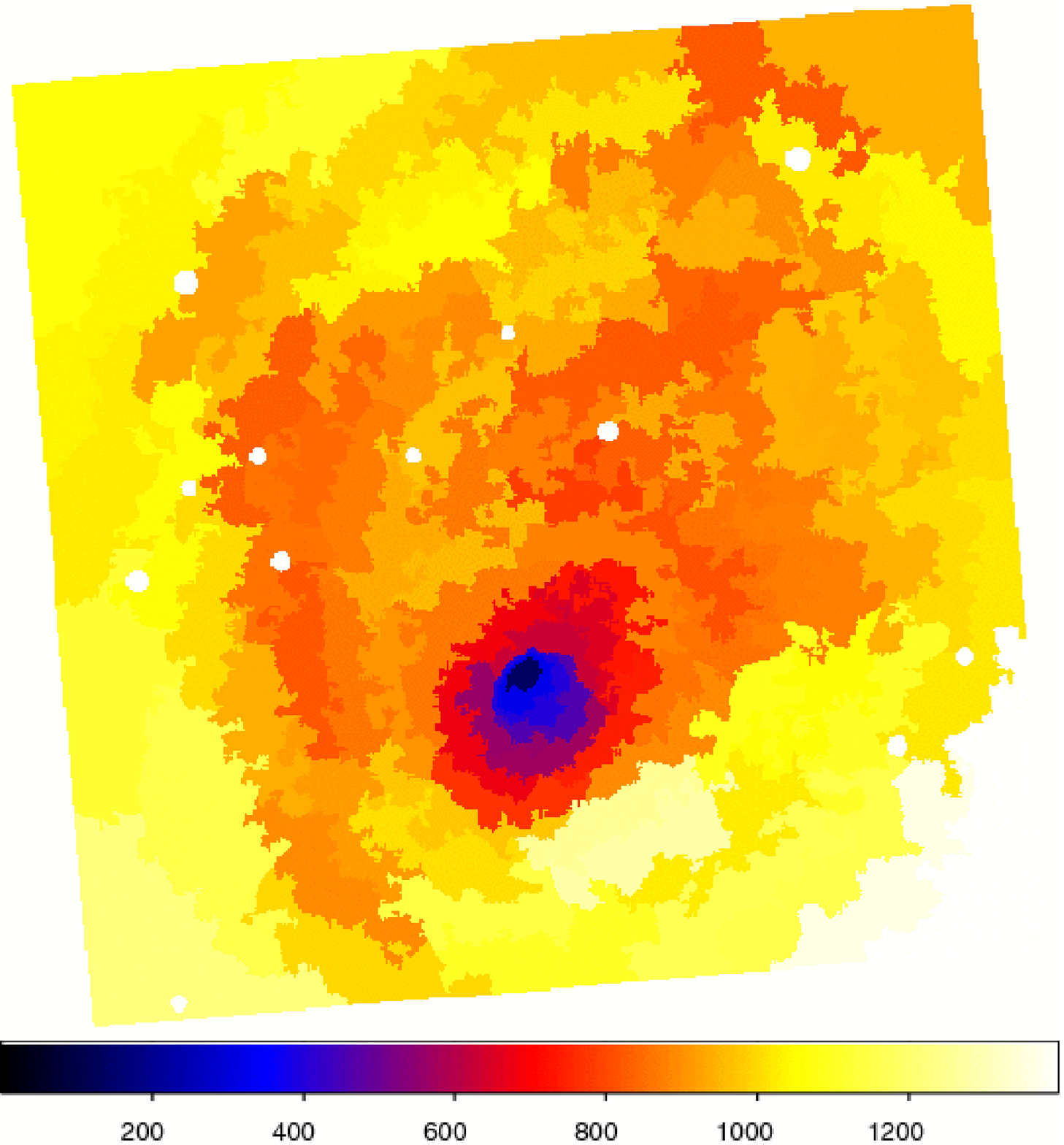}}
\hspace{0.9cm}
\scalebox{0.43}{\includegraphics{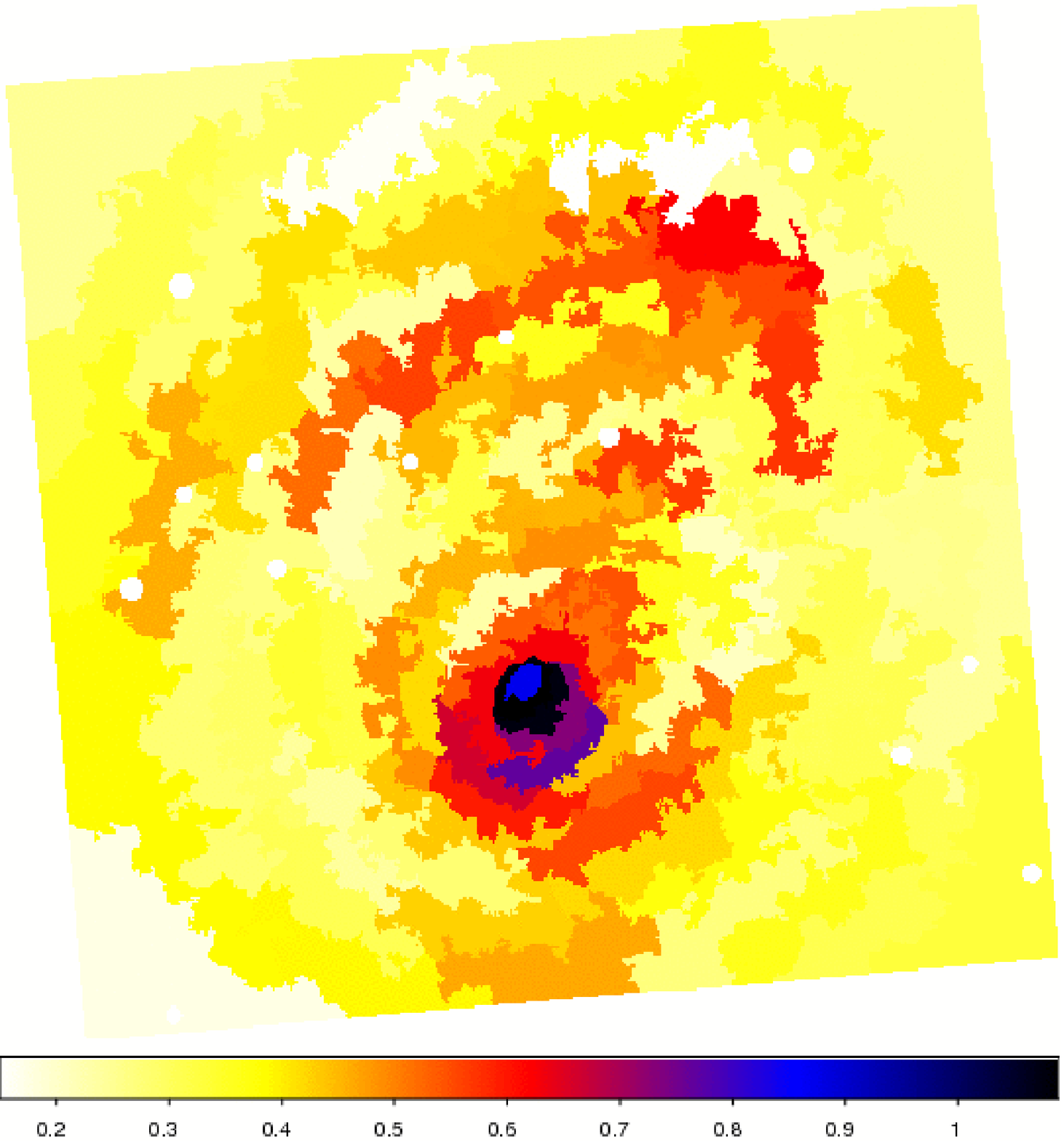}}
\caption{Temperature map (upper left; in keV), pressure map (upper
right; in keV cm$^{-5/2}$ arcsec$^{-1}$), entropy map (lower left; in
keV cm$^{5/3}$ arcsec$^{2/3}$), and metallicity map (lower right; in
solar) for the field of view covered by ACIS Chip 7 
(see Fig. \ref{fig:sb}a).  Individual
regions follow lines of constant surface brightness, while maintaining
a constant signal to noise threshold ($\sim$10,000 net counts per
region). This results in statistical uncertainties on the temperature
and metallicity of $\sim$10 per cent or better. The electron 
number density is approximated
as $\sqrt{K/A}$, where K is the {\small MEKAL} normalization and A 
is the projected area.}
\label{fig:lss}
\end{figure*}

\section{Results}

\subsection{Imaging analysis}
\label{section:image}

Background subtracted, flat-fielded images in the $0.6-7.0$ keV band were
extracted on a $0.984\times0.984$ arcsec$^2$ pixel scale.  Fig.
\ref{fig:sb}a shows the X-ray image for the region spanning the
central ACIS chip (approximately $8\times8$ arcmin$^{2}$). 
The peak of the X-ray emission lies at coordinates
$\alpha(2000)=17^{\rm h}12^{\rm m}27.64^{\rm s}, 
\delta(2000)=-23^{\circ}22^{\rm m}07.0^{\rm s}$.  This peak is embedded
in an extended X-ray bright halo, which has a comet-like tail extending to
the north/northwest.  
A clear surface brightness discontinuity or `cold front' is observed
at a radius of
40 kpc to the south of the X-ray surface brightness peak (first noted
by Ascasibar \& Markevitch 2006).
The overall X-ray morphology suggests that the
cluster core is traveling approximately north to south, along a
position angle of $\sim160$ degrees, with ram-pressure leading to the
formation of the comet-like tail behind it.

Fig. \ref{fig:sb}b shows the central $1.6\times1.6$ arcmin$^2$ region of
the cluster. Two additional sharp surface brightness fronts
(also noted by Ascasibar \& Markevitch 2006), are observed
at radii of $\sim 3$
and $\sim 10$ arcsec to the west and northeast of the
X-ray peak, respectively.  These fronts appear to be sites of
significant shearing and
suggest significant swirling and sloshing within the inner, X-ray
bright cluster core.

Fig. \ref{fig:optical} shows the co-added H, J, K band 2MASS
image for the same, central $1.6\times1.6$ arcmin$^2$ region. The
dominant cluster galaxy is seen near the center of the field.  The
green circle in the figure marks the location of the X-ray peak; the
radius of the circle provides a conservative estimate of the astrometric
uncertainty in the X-ray astrometry ($\sim 1$ arcsec).
As was also reported by P\'erez-Torres \etal (2009), 
the optical centroid is
offset to the south of the X-ray peak by 4 arcsec (2 kpc).
This offset presumably reflects the  impact of ram pressure in 
slowing down the X-ray emitting gas, and the essentially collisionless
nature of the stars (and dark matter) associated with the galaxy, which
carry on ahead.

Fig. \ref{fig:radio} shows the 1.4 GHz NVSS image 
(Condon \etal 1998; the resolution is approximately
$45\times45$ arcsec$^2$; FWHM) for the same field of view
as Fig. \ref{fig:sb}a.  A bright, 
central radio source is observed at a position approximately coincident with
the position of the central galaxy.  A low surface brightness radio feature
extends towards the northwest, 
much like the comet-like trail in X-rays, although the signal-to-noise
ratio is low.
Higher-resolution and deeper multi-frequency radio data are required to 
determine the origin of this feature as well as the detailed relation
to the complex central X-ray emission.

The X-ray surface brightness profile for the cluster about the X-ray
peak is shown in Fig. \ref{fig:sbprof}. The profile is very sharp
and no clear central flattening is observed. The profile can be approximated
by a power law model, $S_0 R^{-\alpha}$, with slope 
$\alpha=0.6112\pm0.0015$, or a $\beta$ model with
$\beta=0.2685\pm0.0002$ and core radius $r_c<0.5$ kpc.
However, neither simple model provides a formally acceptable fit to the data.

\subsection{Thermodynamic mapping}

\subsubsection{Large scale properties}

Fig. \ref{fig:lss} shows the large scale thermodynamic structure of
the cluster, on the same spatial scale as Fig. \ref{fig:sb}a.  From
top to bottom, left to right, we show temperature, pressure, entropy
and metallicity maps.  These quantities show significant morphological
correspondence with the X-ray surface brightness image.  The X-ray bright
core contains relatively cool, low entropy gas. This gas has a
high metallicity. Beyond the central 30kpc radius, the
temperature map is relatively isothermal.

Beyond the inner 30 kpc region, no clear gradients in metallicity are
observed.  However, a large, metal-rich ridge spans position angles of
-50 to 45 degree, at a radius of $\sim$100 kpc to the north of the
cluster core. This ridge consists of 5 independent, coherent, metal
rich regions.  
The mean metallicity of the 5 regions is $0.55\pm0.04$ solar, which
is approximately 1.8 times greater than the mean value of $0.30\pm0.03$ solar
determined for the 9 adjacent regions.
A second ridge of enhanced metallicity appears to
partially encircle the cluster core at a radius of $\sim40$ kpc
spanning position angles of
-55 to 270 degrees at a radius $\sim$40 kpc. To the south, 
where the metallicity of the ridge is highest, it is approximately 
coincident with the location of the southern cold front. This southern
ring-like ridge contains 10 coherent, independent regions of measured
high metallicity. The mean metallicity of these 10 regions is $0.51\pm0.04$
solar, which is approximately 1.5 times greater than the mean value of
$0.34\pm0.04$ solar determined from the 7 adjacent regions at larger
radius.

\begin{figure}
\hbox{
\hspace{0.0cm}
\scalebox{0.43}{\includegraphics{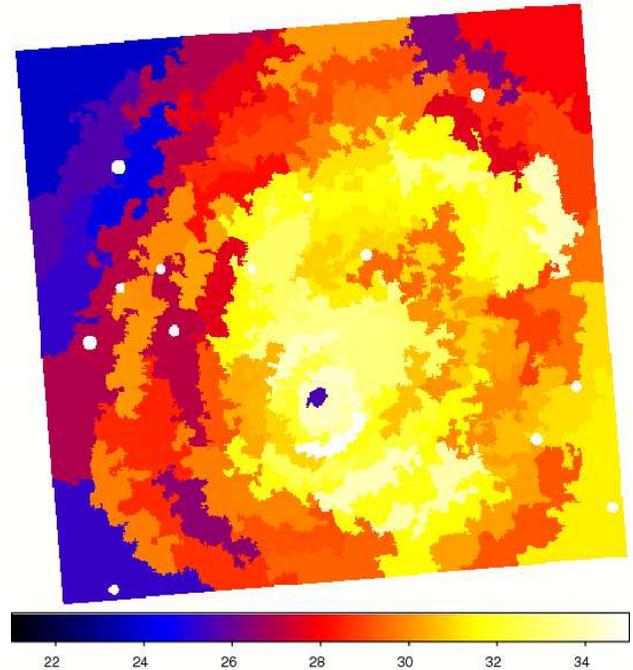}}
}
\caption{Map of the measured equivalent, line-of-sight absorbing
hydrogen column density (in units of
$10^{20}$ atom cm$^{-2}$) for the region covered by ACIS chip 7 (same
region as Fig. \ref{fig:sb}a).
Other details as for Fig. \ref{fig:lss}. The statistical uncertainties
in the measurements are at the  
$\sim5$ per cent level.
}
\label{fig:nhz}
\end{figure}

In the above analysis, the absorption was linked to vary in unison
across all regions.  We have also investigated other absorption
models.\footnote{ The qualitative structures of the thermodynamic maps
are unchanged using different absorption models.  We note, however,
that the exact value of the temperature and quantities derived thereof
vary by up to 10 per cent, depending on the exact value of the column
density.}  When allowing the absorption to be independently free in
each region, we obtain the map shown in Fig. \ref{fig:nhz}.  A trend
towards higher absorption is observed in the cluster center, similar
to Abell 478 (Allen \etal 1993; Sun \etal 2003).  The innermost bin exhibits
somewhat lower apparent absorption, although this may be an
artifact due to the complex thermal structure of the X-ray
emission from this region.  A coherent structure of excess absorption
appears to cross through the cluster, with the western side of the
field being more absorbed than the east.

\subsubsection{Properties of  the inner cluster core}

\begin{figure*}
\hspace{0.0cm}
\scalebox{0.43}{\includegraphics{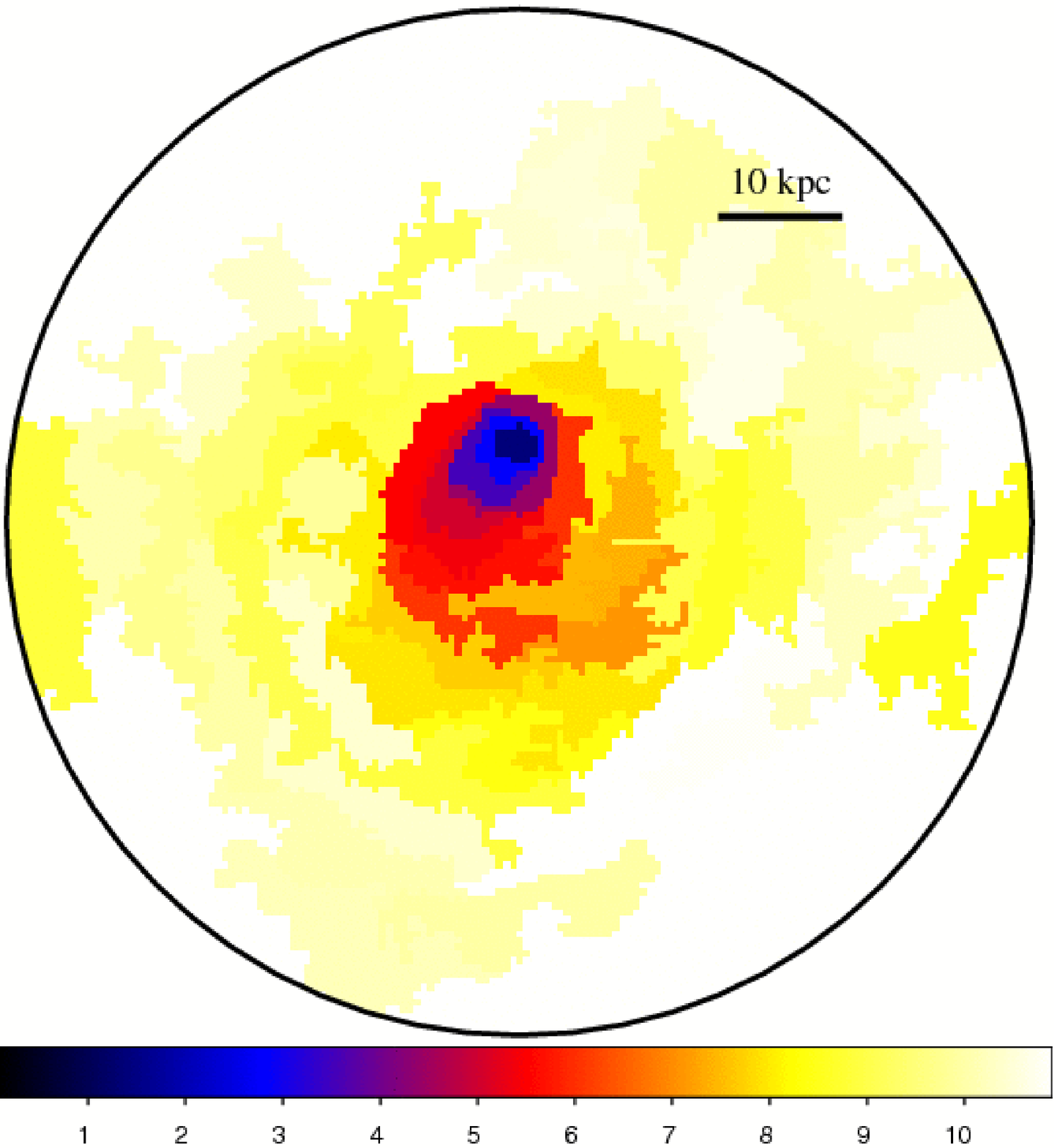}}
\hspace{0.9cm}
\scalebox{0.43}{\includegraphics{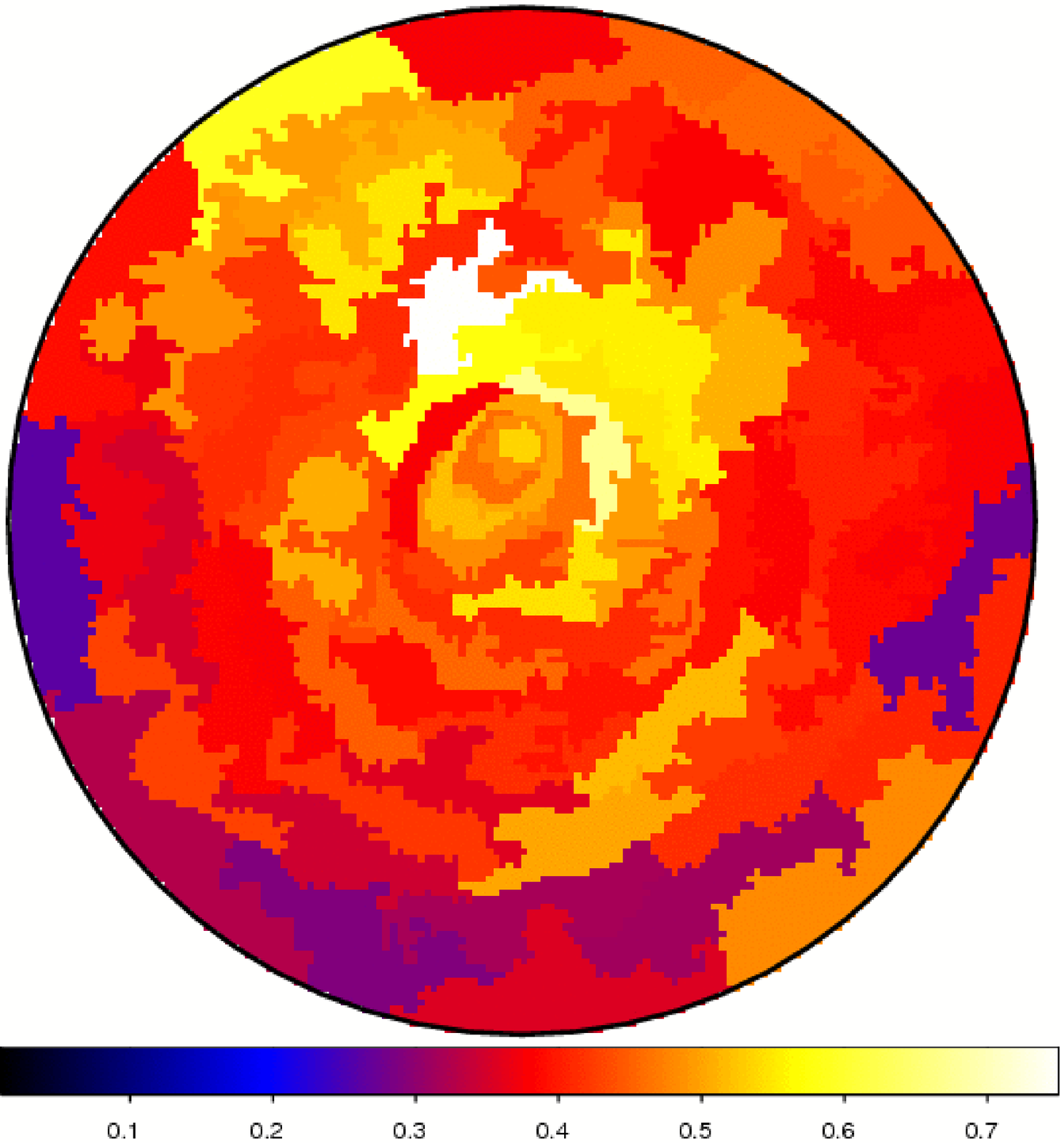}}\\
\hspace{0.0cm}
\scalebox{0.43}{\includegraphics{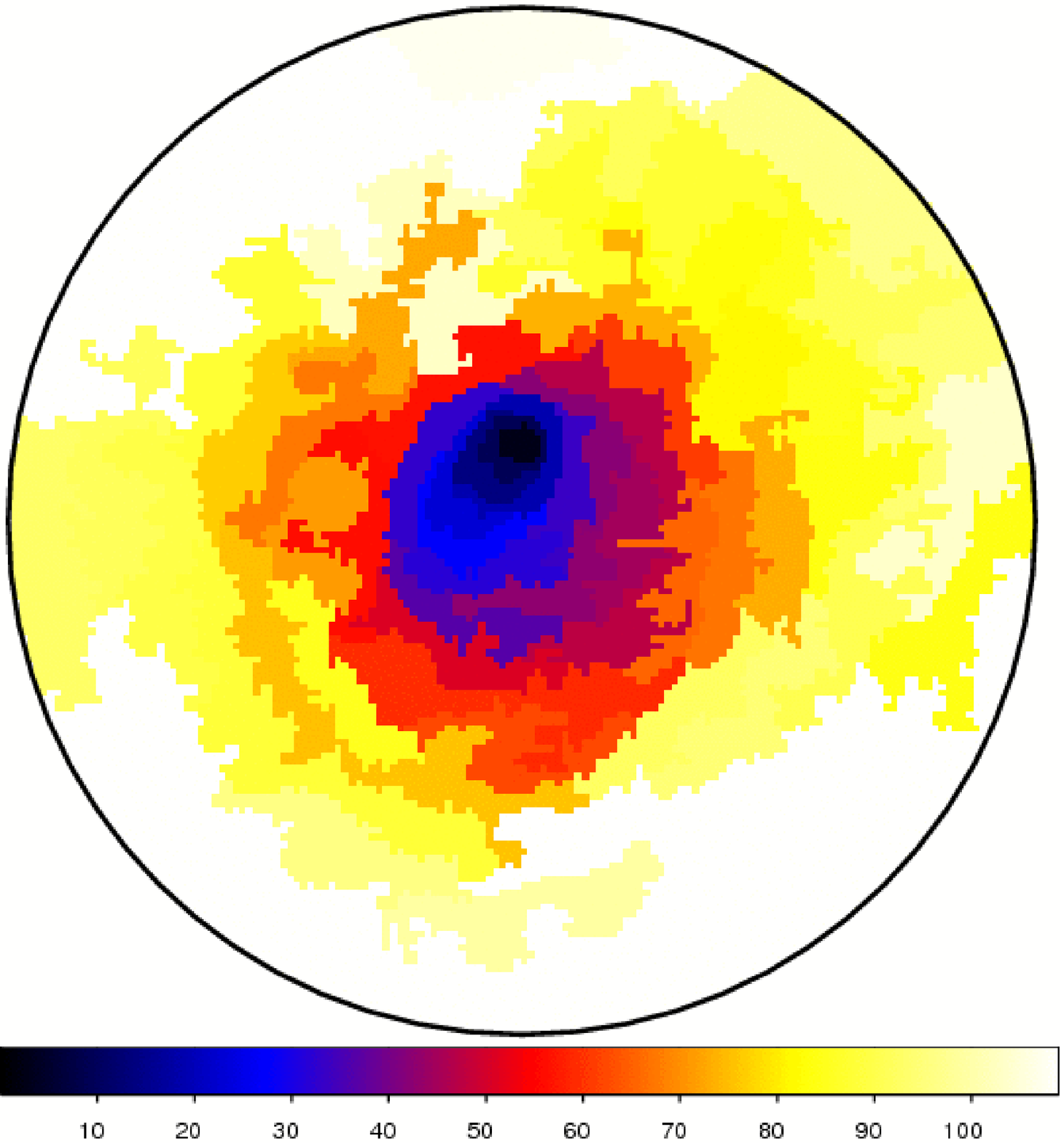}}
\hspace{0.9cm}
\scalebox{0.43}{\includegraphics{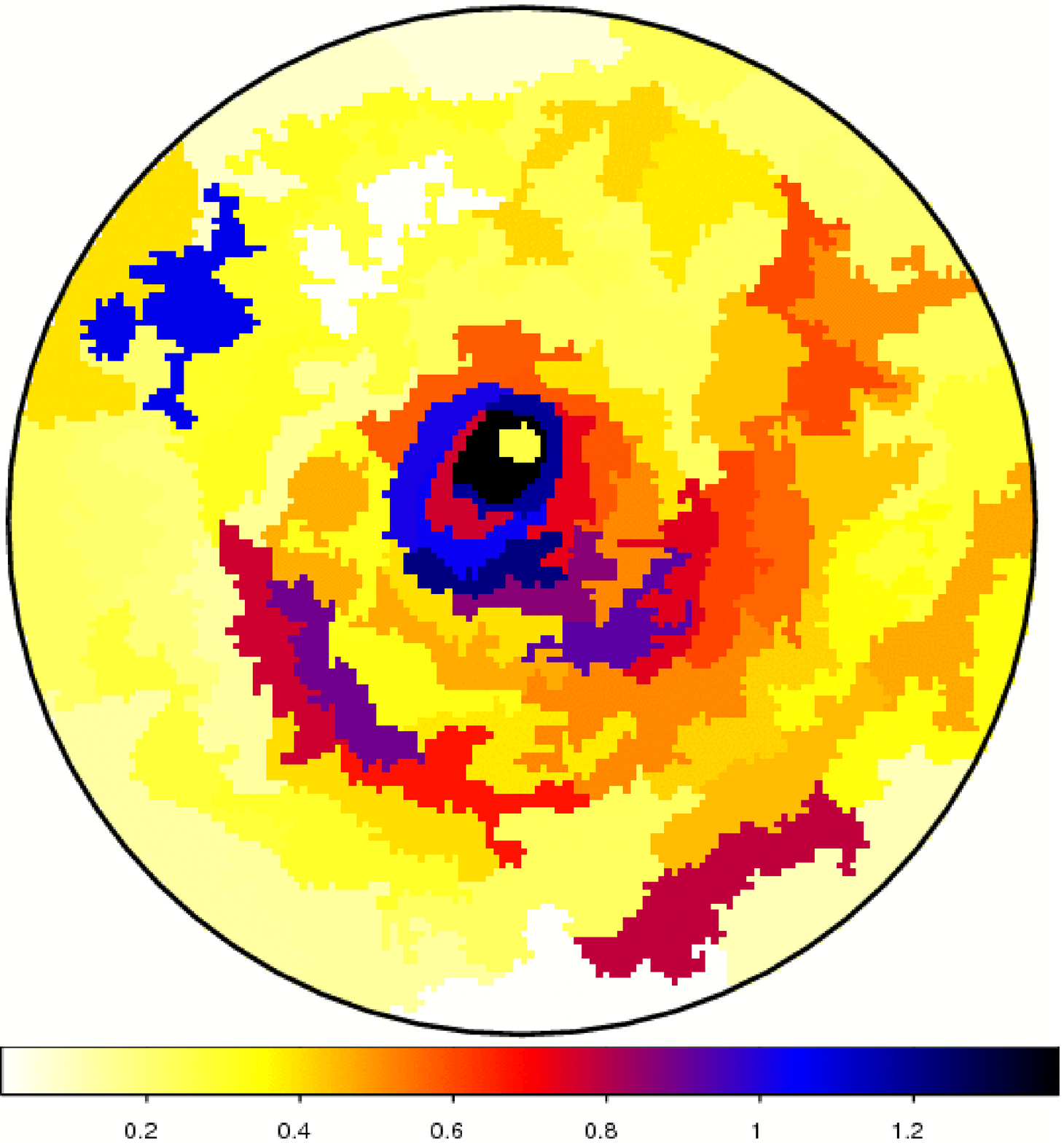}}
\caption{Temperature (upper left, in keV), pressure (upper right, in
keV cm$^{-3}$), entropy (lower left, in keV cm$^{2}$), and metallicity
(lower right, solar) maps for the central $\sim$40 kpc radius of the
Ophiuchus Cluster.  Regions maintain a constant signal to noise
threshold ($\sim$3,000 net counts per region), resulting in 
statistical uncertainties of $\sim$5 per cent in temperature, for
$kT<5$ keV.  Here, we approximate the volume of each 
emission region as
$\frac{2}{3}\,A\,l$, where A is the projected area and $l$ is the line
of sight distance; $l=\sqrt{r_{max}^2-r_{min}^2}$. $r_{max}$ and
$r_{min}$ are the maximum and minimum distances for all points in a
region, from the X-ray surface brightness peak (see Henry \etal
2004, Mahdavi \etal 2005). The number density is calculated from
the {\small MEKAL} normalization.
}
\label{fig:zoom}
\end{figure*}

Fig. \ref{fig:zoom} shows the temperature, pressure, entropy and
metallicity maps for the inner $\sim40$ kpc radius region.
The maps reveal steeply declining temperature and entropy as one moves
inward toward the X-ray peak.  The morphology of the lowest
temperature gas traces a swirl structure also seen in the surface
brightness map (Fig \ref{fig:sb}b).  Interestingly, the X-ray
brightness peak does not coincide with the pressure maximum: the
latter is located to the north of the brightness peak, and trails the
inferred southward motion.

The central region of the cluster shows an enormous range of
temperature, from less than 1 keV to greater than 10 keV. This is only
the second hot, massive cluster to show sub 1.0 keV gas in spectral
maps (see also the study of Abell 2204 by Sanders \etal 2009).  If the
cluster were at significantly higher redshift, this coolest gas would
not be spatially resolved.

\begin{figure*}
\hspace{0.0cm}
\scalebox{0.47}{\includegraphics[angle=270]{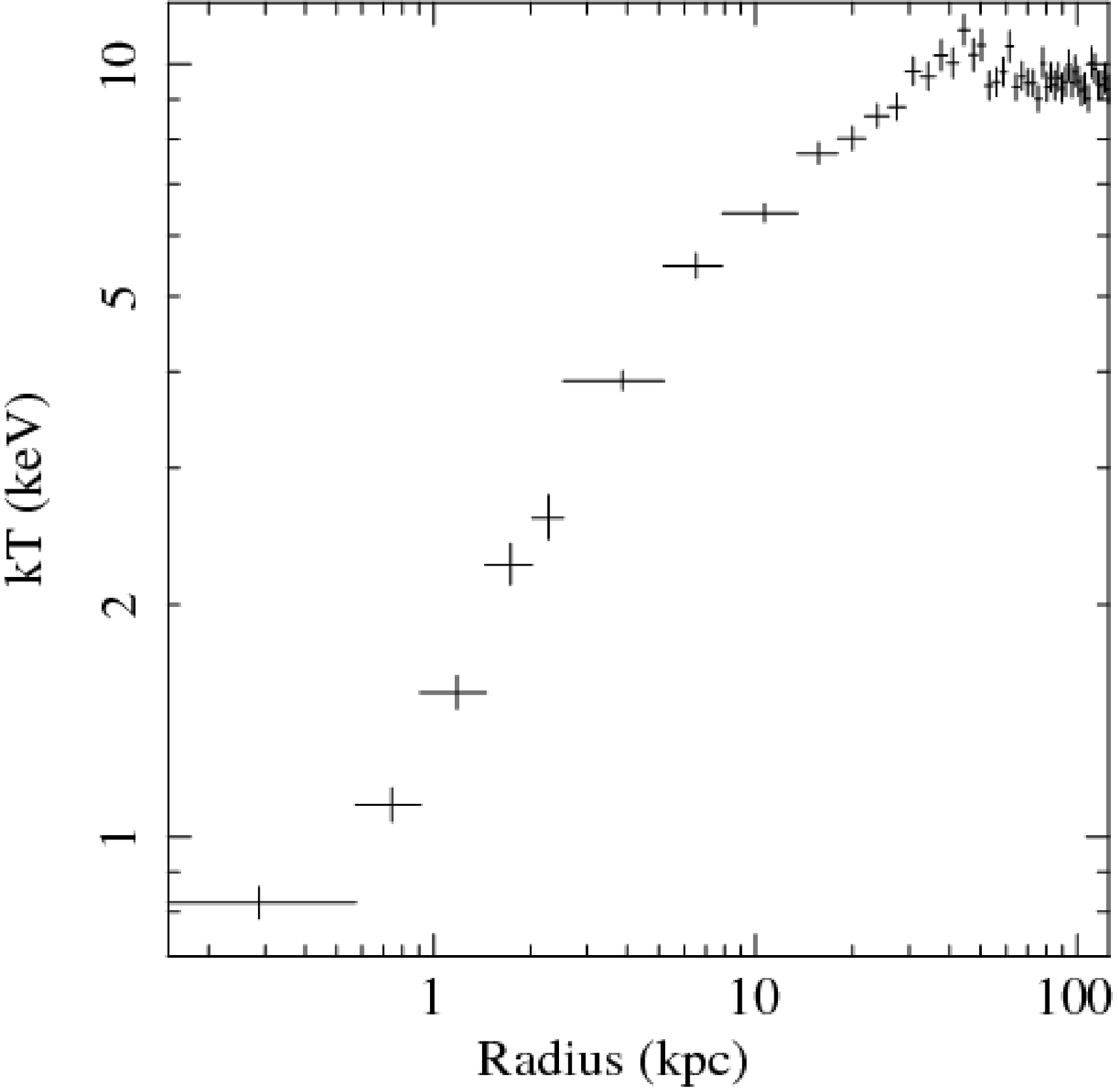}}
\hspace{0.4cm}
\scalebox{0.47}{\includegraphics[angle=270]{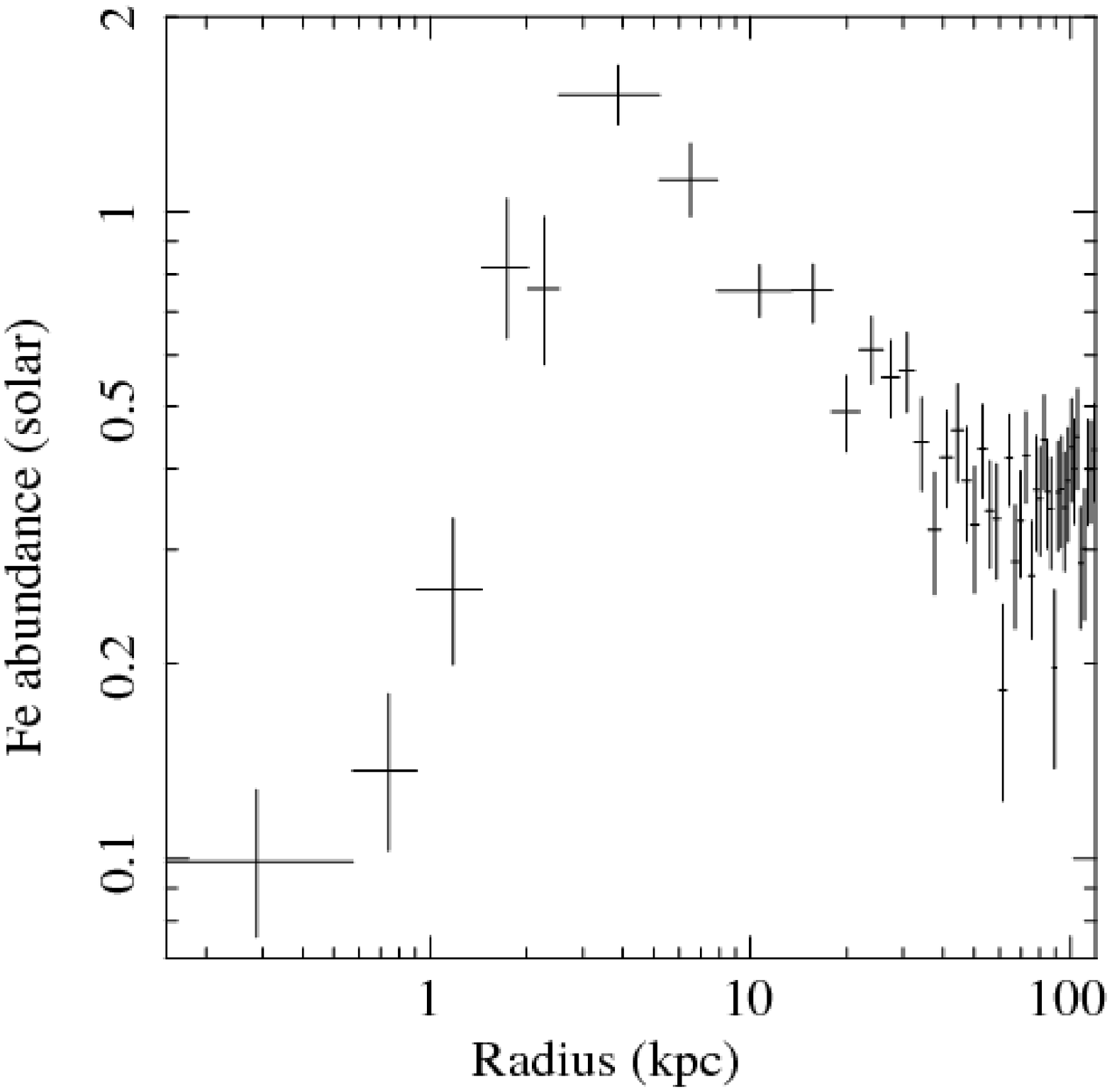}}
\caption{Azimuthally-averaged, projected temperature (left; in keV)
and Fe abundance (right; in solar units) profiles, centered on the X-ray peak. 
The temperature profile
shows a strong central gradient within 30 kpc.  The central abundance
is low, rises to a maximum value $Z=1.5^{+0.17}_{-0.15}$ by 5 kpc,
before declining again to a value $Z\sim0.35$ at 30 kpc. For radii
$r>50$ kpc, the projected temperature and metallicity remain
approximately constant.}
\label{fig:proj}
\end{figure*}

\begin{figure*}
\hspace{-0.1cm}
\scalebox{0.33}{\includegraphics[angle=270]{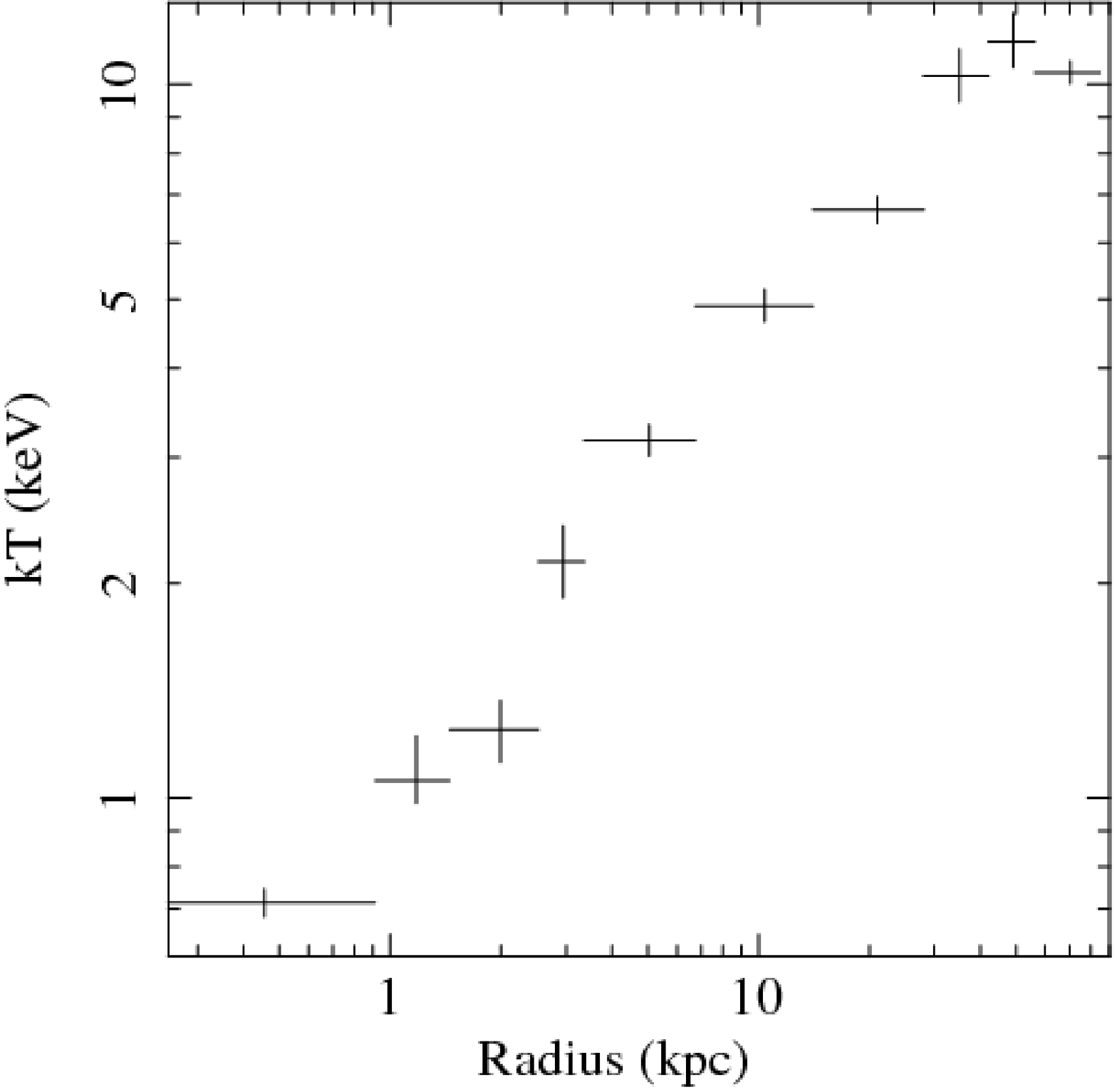}}
\hspace{0.0cm}
\scalebox{0.33}{\includegraphics[angle=270]{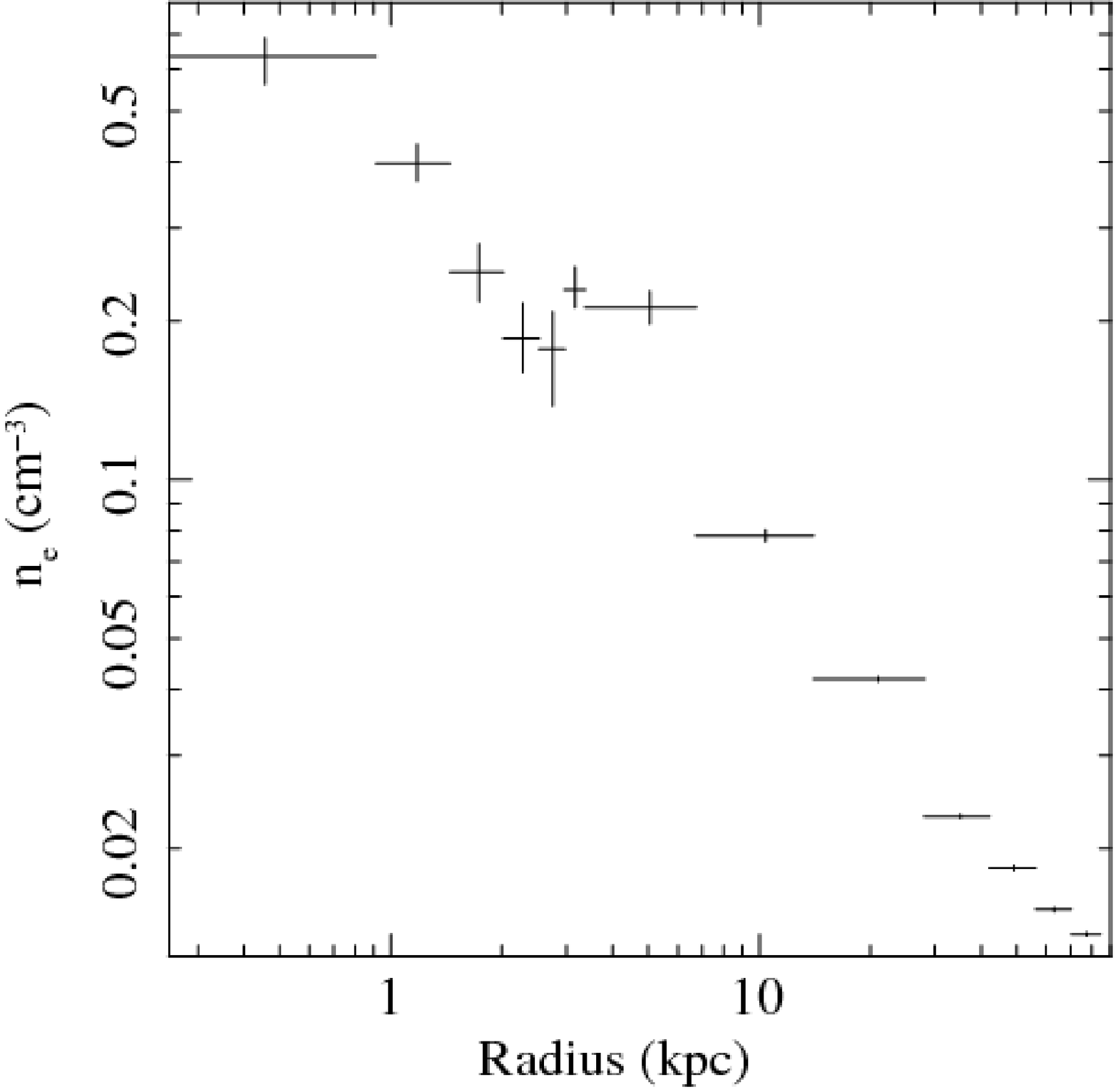}}
\hspace{0.0cm}
\scalebox{0.33}{\includegraphics[angle=270]{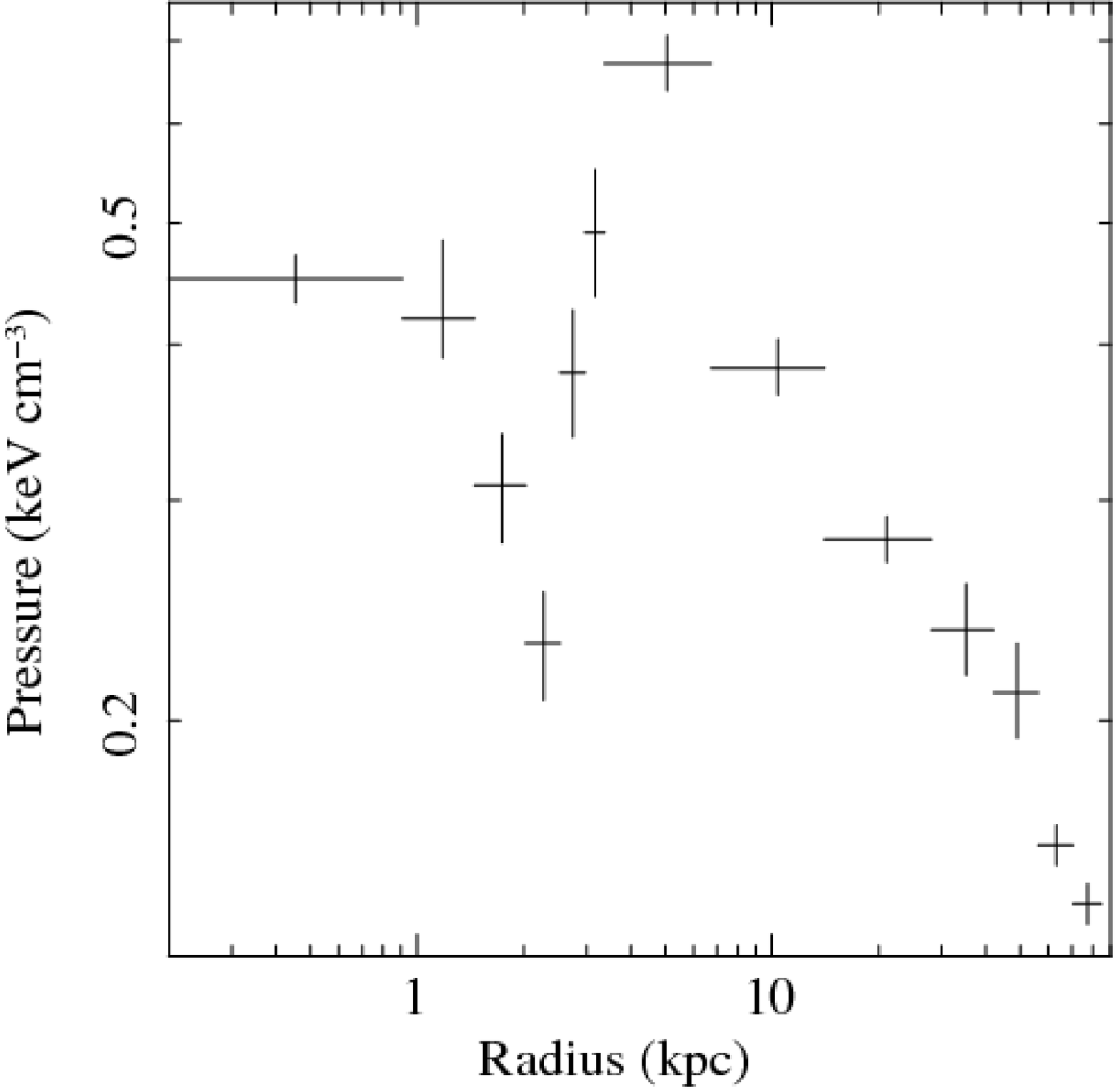}}\\
\vspace{0.5cm}
\hspace{-0.1cm}
\scalebox{0.33}{\includegraphics[angle=270]{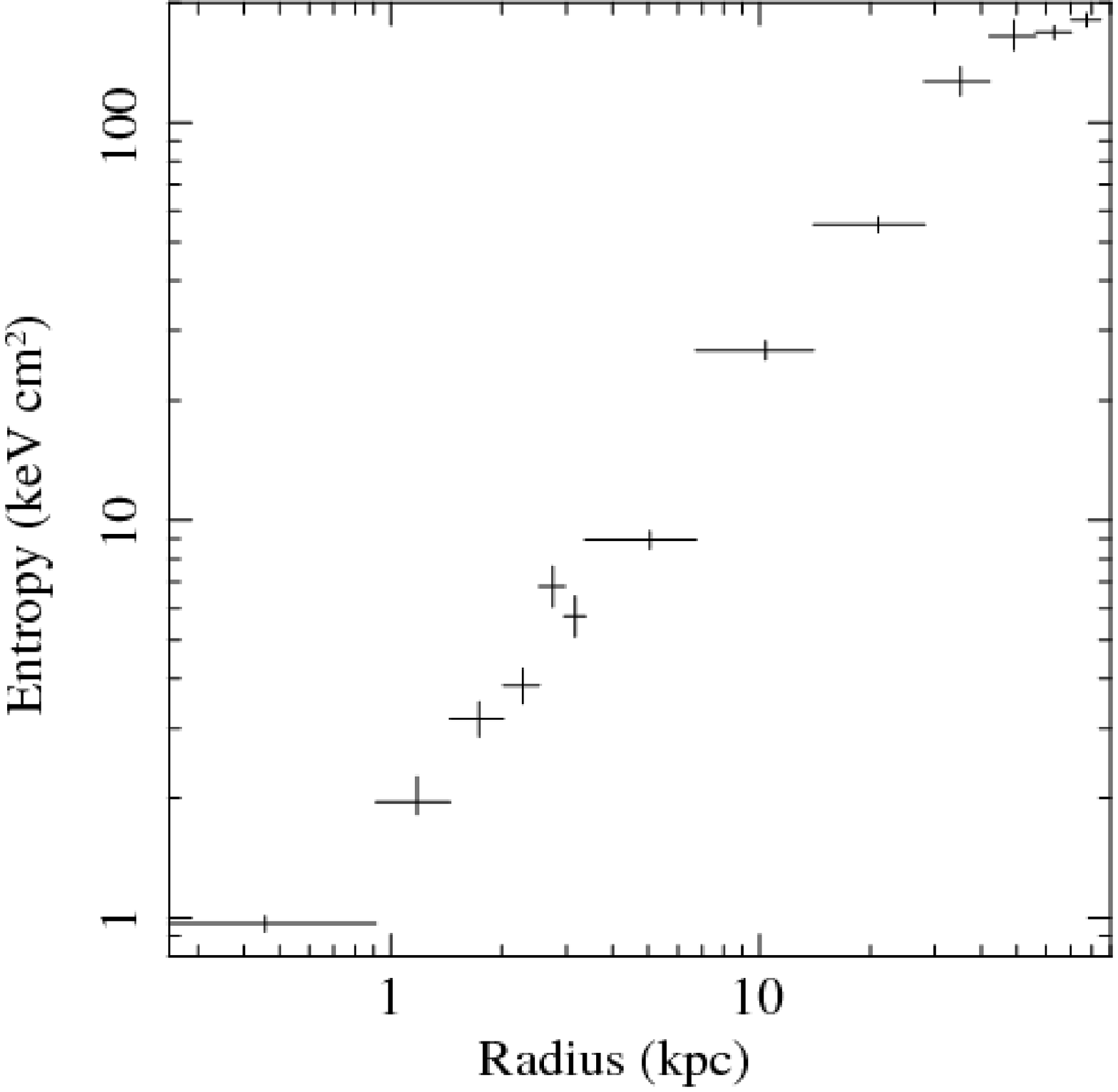}}
\hspace{0.0cm}
\scalebox{0.33}{\includegraphics[angle=270]{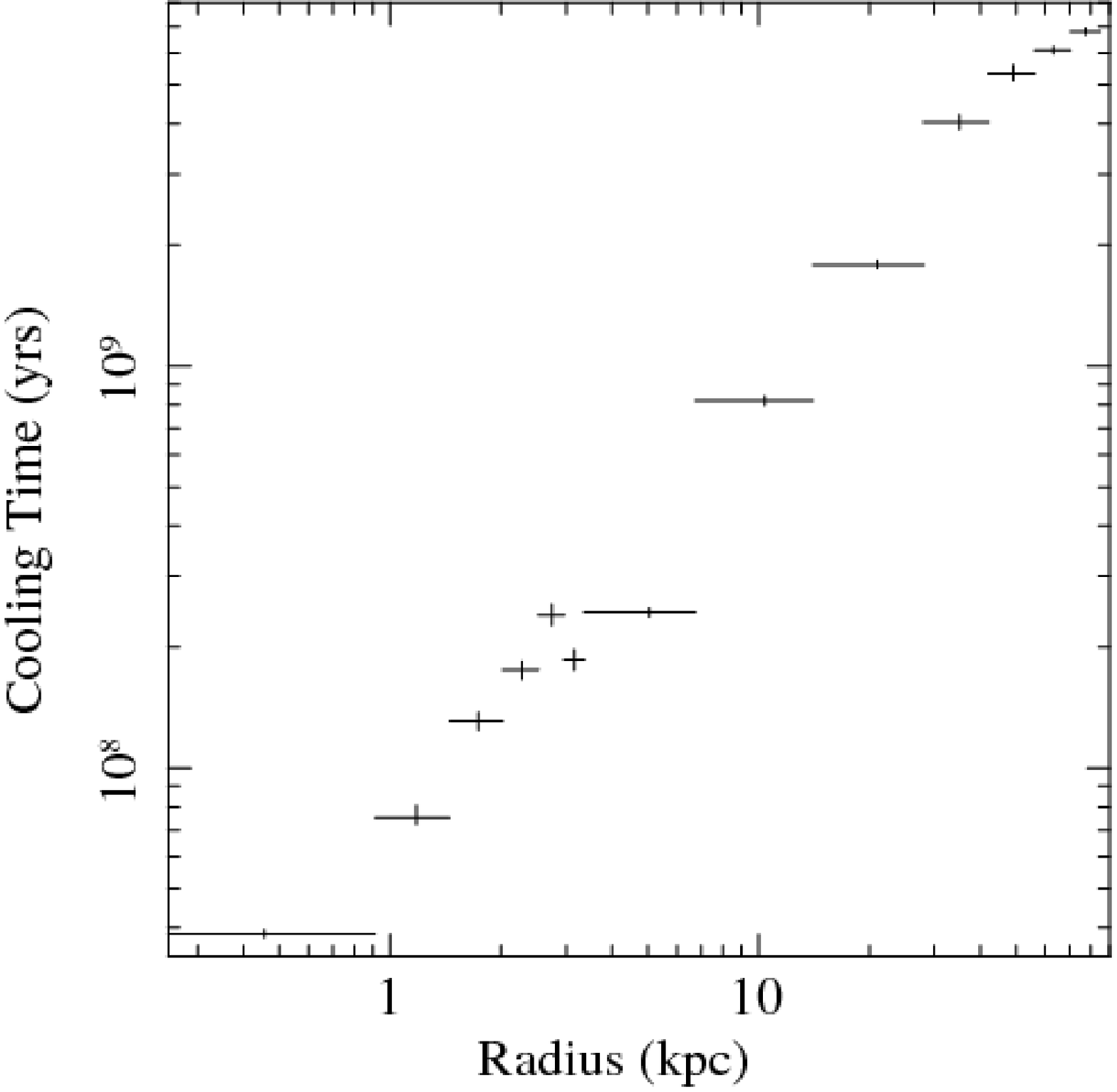}}
\hspace{0.0cm}
\scalebox{0.33}{\includegraphics[angle=270]{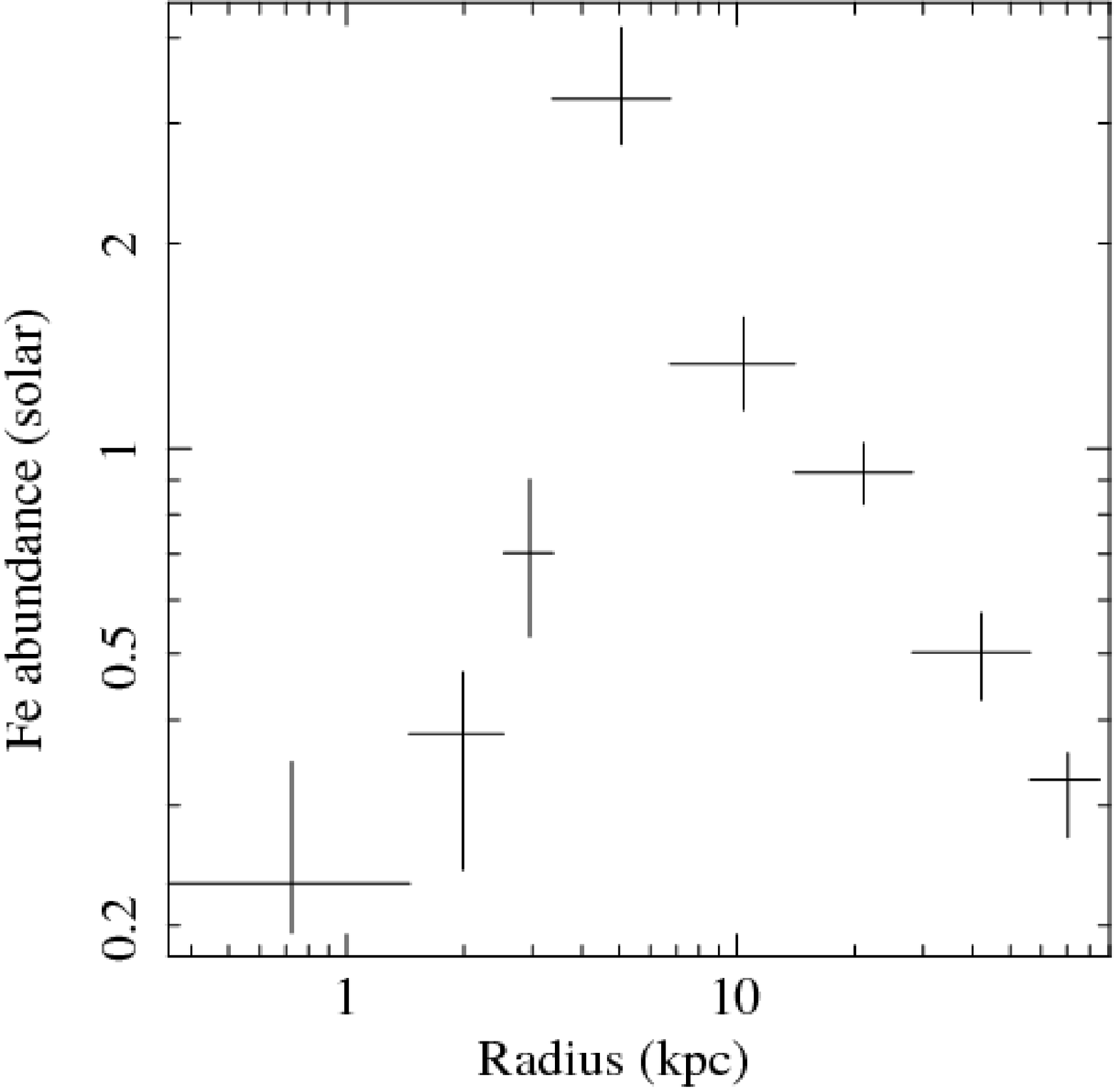}}
\caption{Deprojected temperature (upper left; in keV), electron density (upper
middle; in cm$^{-3}$), pressure (upper right; in keV
cm$^{-3}$), entropy (lower left; keV cm$^2$), cooling time (lower
middle; in years), and Fe abundance (lower right; with respect to
solar) profiles.  Profiles are centered on the X-ray peak and assume
spherical symmetry.  Significant asymmetries in the cluster are
present, although the results obtained from deprojecting partial
angular sectors of the cluster are broadly consistent.
We caution that departures from spherical symmetry will have some effect
on the electron density, pressure, entropy, and cooling time profiles.
}
\label{fig:deproj}
\end{figure*}

\subsection{Azimuthally-averaged, radial profiles}
\label{section:rad}

Fig. \ref{fig:proj} shows the azimuthally-averaged, projected
temperature and metallicity profiles for the cluster, centered on the 
X-ray peak. 
The profiles reveal a steep central temperature gradient within the
central 30 kpc radius.  Beyond $r\sim50$ kpc, the cluster appears
remarkably isothermal.  The same central 30 kpc also exhibits a complex
metallicity profile, which first rises steeply 
with increasing radius, from 0.1 solar within 0.5 kpc to 
almost twice solar at $3-5$ kpc.
Between $5<r<50$ kpc, the metallicity declines steadily before becoming
approximately constant for $r\approxgt50$ kpc, out to the edge of the field.

Fig. \ref{fig:deproj} shows the deprojected temperature, density, pressure,
entropy, cooling time and metallicity profiles, determined under the
assumption of spherical symmetry.
We caution that departures from spherical symmetry will have some effect
on the electron density, pressure, entropy, and cooling time profiles.
The deprojected profiles confirm the presence of a strong, central
temperature gradient, which rises with increasing radius by
over a factor of 10 within the central 30 kpc. The minimum
temperature, measured within the central 1 kpc, is $\sim$0.7 keV, 
consistent with that of large,
elliptical galaxies. Indeed, it may be that this coolest gas was
originally the interstellar medium of the dominant cluster galaxy,
which has been stripped by ram pressure.  By 30 kpc, the temperature
has risen to 10 keV. This is the strongest central temperature
gradient observed in any cluster to date.

The central cooling time is very short, dropping below $4\times10^7$ yrs.
The central metallicity gradient is even more pronounced in the deprojected
profiles,
with $Z \sim 0.25$ solar within 1 kpc, a peak at $Z\sim3$
solar at $r\sim5$ kpc, and dropping again at larger radii.  
We caution, however, that the low central
metallicity may in part be an artifact due to complex temperature 
structure in the innermost regions 
(see Werner \etal 2008 and references within).
Beyond $r\sim50$ kpc, the temperature and metallicity remain
approximately constant at $kT\sim10$ keV and $Z\sim0.35$ solar,
respectively, to the edge of the field of view.

Within 5 kpc of the X-ray peak, the deprojected pressure profile
declines.  The maximum pressure is observed 5 kpc from the center.
Our thermodynamic maps (Fig. \ref{fig:zoom}b) show that the pressure
peak lies to the north of the brightness peak.  Extrapolating the
pressure profile for $r>5$ kpc inward to the core, we estimate that
the innermost regions are under-pressured by $\sim60$ per cent.  This
argues for substantial non-thermal pressure support in these regions,
which is probably dominated by gas motions, as suggested by the
multiple brightness fronts and X-ray-optical displacement seen in
Fig. \ref{fig:optical}. 
Turbulent motions at the level of $\sim500$ km s$^{-1}$ could in
principle explain the thermal pressure deficit.
However, it is also possible that non-thermal 
particles contribute to the total pressure support in the core.
Higher resolution radio data are required to explore this possibility.

\subsection{Analysis of the southern cold front}

A cool, dense core moving through a diffuse hot, gaseous halo will
experience ram pressure.  In detail, the pressure ratio at the cold
front leading the core is related to the core
velocity by (Landau \& Lifshitz 1959; see also
Vikhlinin \etal 2001)
\begin{equation}
\frac{p_0}{p_1}=\left(1+\frac{\gamma-1}{2}M^2\right)^{\frac{\gamma}{\gamma-1}}, M\leq1
\end{equation}
\begin{equation}
\frac{p_0}{p_1}=\left(\frac{\gamma+1}{2}\right)^{\frac{\gamma+1}{\gamma-1}}M^2\left[\gamma-\frac{\gamma-1}{2M^2}\right]^{\frac{-1}{\gamma-1}}, M>1,
\end{equation}
where $M=v/c$ is the Mach number of the core through 
the hot diffuse gas, and $\gamma=5/3$
is the adiabatic index. Here, index 1 refers to the ambient, hot
gas and index 0 to the moving, cool core.  

We have applied these calculations to the southern cold front
at $r\sim40$ kpc, which has a relatively simple geometry and
appears to lead the overall motion of the cool core.  
For this analysis we examine the deprojected pressure profile in a 110 degree
slice, centered on the center of curvature of the front. Accounting for the 
underlying ambient pressure profile, measured in other directions, we 
determine a pressure ratio at the southern cold front of 
$p_0/p_1=1.3\pm0.2$.
This corresponds to a Mach number of
$0.58\pm0.10$ and, for an ambient temperature
kT$=10.6\pm0.3$ keV, implies a velocity of
$1000\pm200$ km s$^{-1}$.

The significant rotational motions that appear to be associated  
with the two other fronts discussed in Section \ref{section:image}
prevent any reliable quantitative determinations of their velocities.

\subsection{A search for possible non-thermal-like emission}
\label{section:nonthermal}

Recently, a detection of a non-thermal X-ray emission component from
the Ophiuchus Cluster was claimed by Eckert \etal (2008)
using data from the $INTEGRAL$ satellite.  
These results are in apparent 
tension with those from {\it Suzaku} and $Swift/BAT$
(see Fujita \etal 2008; 
Ajello \etal 2009).\footnote{A re-analysis of $INTEGRAL$ data, also
using {\it XMM-Newton} data to constrain
the thermal emission at lower energies (Nevalainen \etal 2009), is 
consistent with the upper limits obtained by Fujita \etal (2008) and Ajello
\etal (2009).}
In general, proper temperature modelling is crucial
in determining the presence, or otherwise, of non-thermal spectral components
in the X-ray band. As shown in Fig. \ref{fig:proj}, a steep
temperature profile exists in the central 30 kpc ($\sim1$ arcmin)
of the cluster and must be accounted for.
Residual 
calibration uncertainties will also
affect the measured temperature and the 
inferred non-thermal emission properties.
For example, a change in the globally-averaged temperature of $\sim$1 keV, 
for a kT$\sim10$ keV
system like the Ophiuchus Cluster,
can change the expected thermal flux above 20 keV by $\sim30$ per cent.
Such effects currently make it difficult to provide 
any definite statements on the presence
or otherwise of possible non-thermal X-ray spectral components in the 
Ophiuchus Cluster.

Despite these difficulties, 
we have searched for non-thermal spectral components in the $Chandra$ data
following the procedures outlined by
Million \& Allen (2009). In detail, 
we have examined the statistical significance of the
improvements to the fits that can be obtained with the inclusion of additional
power-law components in the spectral modelling. The normalization of 
the power-law component in each spectral region was a free parameter, while
the photon index was fixed to $\Gamma=2.0$. The statistical signal
from all regions was combined to determine the final result on
the presence of possible power-law emission.

We find no statistical signal of 
non-thermal-like components in the $Chandra$ data, once
the complex, underlying temperature structure revealed by our analysis is
taken into account.
Allowing the column density in each region to be fit
independently, we obtain a 
90 per cent confidence upper limit on the flux of any $\Gamma=2.0$
power-law component from the central $8\times8$ arcmin$^2$ region covered by
the ACIS-S chip 7 (excluding only the central 5 kpc region where the 
temperature drops precipitously and projection effects will be large)
of $2.2\times10^{-11}$ ergs s$^{-1}$ cm$^{-2}$ in the
$0.6-7.0$ keV band.\footnote{Our upper limits are determined from the 
sums of the power-law normalizations for the 95 regions with each region having
an MCMC chain of length 10$^4$ samples after correcting for burn in.} 
This upper limit is approximately 10 per cent of the 
measured total flux in the same region. Our results are
consistent with those of P\'erez-Torres \etal (2009), Ajello \etal (2009),
and Fujita \etal (2008).
Our upper limits are at a similar level to the reported detection of 
non-thermal flux by Nevalainen \etal (2009) for the central 7
arcmin radius region.
We note that our result depends slightly upon assumptions
regarding the Galactic absorption in the direction of the cluster.
Fixing the column density for all regions to $N_H=3\times10^{21}$ atom 
cm$^{-2}$ (the average value determined for all regions), our upper limit is 
increased to $2.7\times10^{-11}$ ergs s$^{-1}$ cm$^{-2}$.
Due to observing energy band and field of view differences, direct comparisons
with $INTEGRAL$ results (Eckert \etal 2008) are 
challenging.
Additional hard X-ray data, for example from the forthcoming {\it NuSTAR}
satellite (Harrison \etal 2005) are required to rule definitively on this issue.

\section{Discussion}

The comet-like X-ray morphology, large-scale thermodynamic maps, and
clear pressure jump at the location of the southern cold front
(Figs. \ref{fig:sb}a, \ref{fig:lss}, \ref{fig:zoom}), 
show that the cool, dense core of the 
cluster is moving rapidly through the large-scale hot, diffuse cluster
medium. The velocity implied by the observed pressure jump
at the southern cold front is $\sim1000$ km s$^{-1}$.  The peculiar
velocity of the cD galaxy of $\sim650 $ km s$^{-1}$ with respect to the 
cluster mean (Wakamatsu \etal 2005; Hasegawa \etal 2000) 
also suggests significant,
line-of-sight motion. 
Such speeds may indicate extreme `sloshing' of the cluster core 
(\eg Markevitch \etal 2001; Reiprich \etal 2001; Tittley \& Henricksen 2005;
Ascasibar \& Markevitch 2006; see also Markevitch \& Vikhlinin 2007 
and references within).

The X-ray emission associated with the cool core is very sharply
peaked. The coolest ($kT \sim 0.7$ keV), lowest-entropy gas lies
within 1 kpc of the brightness peak. This material has a very short
cooling time, with $t_{\rm cool} \sim4\times10^7$yr, and plausibly
represents the stripped interstellar medium of the 
dominant cluster galaxy.

The high relative velocity of the cool core argues that it 
was displaced from the base of the cluster potential well, presumably  by recent
merger activity.  This is consistent with the slightly offset location of the
X-ray peak with respect to the X-ray centroid at larger scales (Arnaud \etal 
1987). 
The case for a recent major merger event is supported by the 
significant substructure in the galaxy velocity histogram.
Given the
environment of the present-day cluster, which lies at the center of a
supercluster and is surrounded by smaller galaxy groups and clusters,
a recent merger event would not be surprising.

The degree to which cool cores are disrupted or destroyed during
cluster mergers is an outstanding question in cluster physics,
with important implications for cosmological studies (\eg
Burns \etal 2008; Mantz \etal 2009a,b; Ebeling \etal 2009).
The comet-like morphology of the cool core, and the remarkably steep
central temperature gradient (rising from $\sim
0.7$ keV to 10 keV within a 30 kpc radius) suggest that its outer
regions have already been stripped by ram pressure, 
due to its rapid motion.
Our data provide a
dramatic, close-up view of the ongoing stripping 
and potential destruction of a cool core in a rich cluster
and have clear implications for the
survivability of such cool cores.  It will be important to examine,
using constrained hydrodynamical simulations, the 
evolutionary path of such a cool core as it continues to be stripped 
and its motion slows.  At this point it is not clear how much of the original 
cool core will remain by the time the system returns to 
equilibrium. 

The X-morphology of the inner 10 kpc region (Fig 1b), the separation
between the X-ray and optical/near-IR brightness peaks (the X-ray
emission trails the cD galaxy) and the offset between the X-ray peak
and thermodynamic pressure maximum (the pressure maximum trails the
X-ray brightness peak), all suggest additional, significant motions $within$
the cool core.  The sharp fronts noted in Fig. 1b suggest that
significant rotational motion is present, and that the coolest, X-ray
brightest gas is moving within the central potential.
Such motion is not surprising, given the separation of X-ray peak from
the stellar and, presumably, dark matter potentials of the cD galaxy
(Fig. 2).\footnote{More dramatic 
offsets between the X-ray emitting gas, and stellar
and dark matter mass, have previously seen in massive, bimodal cluster
mergers such as the Bullet Cluster (Clowe \etal 2006; Bradac \etal
2006) and MACS\,J0025.4-1222 (Bradac \etal 2008).} The sharp fronts
seen in Fig 1b are likely to be sites of shearing instabilities, which
could lead to rapid mixing of the coolest gas with its surrounding
environment.

The ridges of enhanced metallicity in the cluster are particularly
interesting. These large-scale, coherent structures are almost twice
as metal rich as the surrounding gas. The prominent metallicity ridge
100 kpc to the north of the X-ray peak, and trailing in its wake, 
appears likely to have been 
stripped from the cool core during its motion.
This ridge, and the partial ring-like ridge of high metallicity
material at a radius of $\sim$40 kpc, may initially have been driven
outward by the central AGN, possibly even during the same outburst.
The proximity of this ring to the southern
cold front argues that this metal rich gas is currently being stripped
from the cool core.  A qualitatively similar high-metallicity ridge is
also seen in the central regions of the Perseus Cluster (Sanders \etal
2005).

Govoni \etal (2009) show that the cD galaxy is surrounded by a
bright, diffuse mini-halo of synchrotron radio emission.  Despite the
fact that a significant non-thermal particle population must be
present to explain the radio emission, our study provides no evidence
for non-thermal X-rays from the central regions of the
cluster. High-resolution radio data for the cluster
core are not currently available.  It will be important to obtain
such data and examine the impact of the cluster dynamics on the
properties of the central radio source, and the interaction between
this source and its surrounding environment.

\section*{Acknowledgments}

We thank Anja von der Linden for help with the analysis of the 2MASS data, 
Glenn Morris for
computational support, and Jeremy Sanders for making the contour
binning algorithm publicly available.
The computational analysis was
carried out using the KIPAC XOC compute cluster at Stanford University
and the Stanford Linear Accelerator Center (SLAC).  
Norbert Werner is supported by the National Aeronautics
and Space Administration through Einstein Postdoctoral Fellowship
Award Number PF8-90056 issued by the Chandra X-ray Observatory
Center, which is operated by the Smithsonian Astrophysical
Observatory for and on behalf of the National Aeronautics and
Space Administration under contract NAS8-03060.
This work was supported in part by the
U.S. Department of Energy under contract number DE-AC02-76SF00515.

\end{document}